\documentclass[10pt,twocolumn,twoside]{IEEEtran}

\usepackage{cite}
\usepackage{graphicx}
\usepackage{psfrag}
\usepackage{amsmath,amssymb}
\usepackage{algorithmic}
\def\reals{{\mathbb{R}}}
\def\complex{{\mathbb{C}}}

\def\ints{{\mathbb{Z}}}

\def\bfZero{\boldsymbol{0}}

\def\bfa{{\boldsymbol{a}}}

\def\bfb{{\boldsymbol{b}}}

\def\bfc{{\boldsymbol{c}}}

\def\bfd{{\boldsymbol{d}}}

\def\bfr{{\boldsymbol{r}}}

\def\bfs{{\boldsymbol{s}}}
\def\bfsh{{\hat{\boldsymbol{s}}}}
\def\bfst{{\tilde{\boldsymbol{s}}}}

\def\bfu{{\boldsymbol{u}}}

\def\bfv{{\boldsymbol{v}}}

\def\bfw{{\boldsymbol{w}}}

\def\bfx{{\boldsymbol{x}}}

\def\bfy{{\boldsymbol{y}}}

\def\bfA{{\boldsymbol{A}}}

\def\bfB{{\boldsymbol{B}}}

\def\bfC{{\boldsymbol{C}}}

\def\bfD{{\boldsymbol{D}}}

\def\bfG{{\boldsymbol{G}}}

\def\bfGt{{\tilde{\boldsymbol{G}}}}

\def\bfH{{\boldsymbol{H}}}

\def\bfI{{\boldsymbol{I}}}

\def\bfM{{\boldsymbol{M}}}

\def\bfP{{\boldsymbol{P}}}

\def\bfQ{{\boldsymbol{Q}}}

\def\bfR{{\boldsymbol{R}}}

\def\bfT{{\boldsymbol{T}}}

\def\bfU{{\boldsymbol{U}}}

\def\bfUt{{\tilde{\boldsymbol{U}}}}

\def\bfV{{\boldsymbol{V}}}

\def\bfW{{\boldsymbol{W}}}

\def\bfX{{\boldsymbol{X}}}
\def\bfXh{{\hat{\boldsymbol{X}}}}

\def\bfY{{\boldsymbol{Y}}}

\def\bfalpha{{\boldsymbol{\alpha}}}

\def\bfbeta{{\boldsymbol{\beta}}}

\def\bfPi{{\boldsymbol{\Pi}}}

\def\bfSigma{{\boldsymbol{\Sigma}}}

\def\bfUpsilon{{\boldsymbol{\Upsilon}}}

\def\Aset{\mathcal{A}}

\def\Bset{\mathcal{B}}

\def\Cset{\mathcal{C}}

\def\Eset{\mathcal{E}}

\def\Fset{\mathcal{F}}

\def\Gset{\mathcal{G}}

\def\Hset{\mathcal{H}}

\def\Nset{\mathcal{N}}

\def\Tset{\mathcal{T}}

\def\Uset{\mathcal{U}}

\def\Xset{\mathcal{X}}

\newcommand{\Diag}{\mathrm{Diag}}

\newcommand{\rank}{\mathrm{rank}}

\def\tr{^{\mathrm T}}
\def\hr{^{\mathrm H}}

\def\fro{{\mathrm F}}

\def\vec{\mathrm{vec}}

\newcommand{\prob}[1]{\mathrm{P}\left(#1\right)}
\newcommand{\tprob}[1]{\mathrm{P}(#1)}
\newcommand{\expt}[2][]{\mathrm{E}_{#1}\left\{ #2 \right\}}

\DeclareMathOperator*{\dotleq}{\overset{.}{\leq}}
\DeclareMathOperator*{\dotgeq}{\overset{.}{\geq}}
\DeclareMathOperator*{\dotle}{\overset{.}{<}}
\DeclareMathOperator*{\dotge}{\overset{.}{>}}
\DeclareMathOperator*{\defeq}{\triangleq}

\newtheorem{theorem}{Theorem}

\newtheorem{lemma}{Lemma}
\newtheorem{definition}{Definition}

\def\ml{\mathrm{ML}}
\def\nt{{n_\mathrm{T}}} 
\def\nr{{n_\mathrm{R}}}

\def\constme{\mathbb{S}}
\def\const{\constme_\eta}
\def\kron{\otimes}

\def\lay{\mathrm{lay}}

\begin{document}
\sloppy
%
\title{Sphere decoding complexity exponent for decoding full rate codes over the quasi-static MIMO channel}

\author{Joakim~Jald\'en and Petros Elia
\thanks{The research leading to these results has received funding from the European Research Council under the European Community's Seventh Framework Programme (FP7/2007-2013) / ERC grant agreement no 228044, and from the Swedish Foundation for Strategic Research (SSF) under the project grant ICA08-0046. P. Elia acknowledges funding by the Mitsubishi RD project Home-eNodeBS. A shortened version of the work is in preparation for submission to the IEEE International Symposium on Information Theory (ISIT-2011).}
\thanks{J. Jald\'en is with the ACCESS Linnaeus Center, Signal Processing Lab, School of Electrical Engineering, KTH - Royal Institute of Technology, Stockholm, Sweden (email: jalden@kth.se)}
\thanks{P. Elia is with the Mobile Communications Department, EURECOM, Sophia Antipolis, France (email: elia@eurecom.fr)}
}

\markboth{Submitted to the IEEE Transactions on Information Theory, February 2011}{Submitted to the IEEE Transactions on Information Theory, February 2011}
\maketitle

\begin{abstract}
In the setting of quasi-static multiple-input multiple-output (MIMO) channels, we consider the high signal-to-noise ratio (SNR) asymptotic complexity required by the sphere decoding (SD) algorithm for decoding a large class of full rate linear space-time codes. With SD complexity having random fluctuations induced by the random channel, noise and codeword realizations, the introduced \emph{SD complexity exponent} manages to concisely describe the computational reserves required by the SD algorithm to achieve arbitrarily close to optimal decoding performance. Bounds and exact expressions for the SD complexity exponent are obtained for the decoding of large families of codes with arbitrary performance characteristics. For the particular example of decoding the recently introduced threaded cyclic division algebra (CDA) based codes -- the only currently known explicit designs that are uniformly optimal with respect to the diversity multiplexing tradeoff (DMT) -- the SD complexity exponent is shown to take a particularly concise form as a non-monotonic function of the multiplexing gain. To date, the SD complexity exponent also describes the minimum known complexity of any decoder that can provably achieve a gap to maximum likelihood (ML) performance which vanishes in the high SNR limit.
\end{abstract}


\section{Introduction}
\label{sec:intro}

The past decade has seen the abundant use of the sphere decoding (SD) algorithm \cite{VB:99,AEV:02,DGC:03,MGD:06} as a tool for facilitating near maximum likelihood (ML) decoding over the coherent delay-limited (or quasi-static) multiple-input multiple-output (MIMO) channel.  The SD algorithm allows for efficient optimal or near optimal decoding of a large number of high rate space-time codes that map constituent constellation symbols linearly in space and time \cite{VB:99}. As the algorithm's computational cost depends on the fading channel, it is generally known that in implementing the SD algorithm, one can tradeoff computational complexity for error performance by selectively choosing when to decode and when not to.  Equivalently, in the presence of constraints on the computational reserves that may be allocated to decoding, the algorithm is faced with the prospect of encountering channel realizations that force it to violate its run-time constraints, thus having to declare decoding outages that inevitably reduce reliability. This naturally raises the intriguing question of how large computational reserves are actually required for near ML performance.

While this question is hard to answer in general, or even ask in a rigorously meaningful way, we show herein that by following \cite{ZT:03} and considering the decoding of sequences of codes in the high signal-to-noise ratio (SNR) limit, not only can the question be made rigorous: It also admits surprisingly simple explicit and general answers. Drawing from the diversity multiplexing tradeoff (DMT) setting which has already been successfully applied to concisely describe the high SNR diversity exponent in the reliability analysis of reduced complexity decoders \cite{KCM:09,TK:10,JE:10}, we introduce the \emph{SD complexity exponent} as a measure of complexity of the SD algorithm. The SD complexity exponent characterizes the decoding complexity in the high SNR limit under the assumption that the code-rate scales with SNR in order to provide a given multiplexing gain. This approach naturally takes into account the dependency of the SD complexity on the codeword density and the codebook size, as well as the SNR and the fading characteristics of the wireless channel. Similar to previous work on the DMT relating the code rate and probability of decoding error, it is seen that also the complexity, although hard to characterize at any finite SNR, has mathematically tractable characterizations in the high SNR asymptote. These characterizations in turn yield valuable insights into the behavior of the algorithm.

The SD algorithm is equivalent to a branch-and-bound search \cite{MGD:06} over a regular tree and like most other works on SD complexity \cite{AEV:02,JO:04,JO:05,MGD:06,SJS:09} we view the number of visited nodes $N$ as the complexity of the algorithm\footnote{In the context of the DMT this is, as we argue later on, equivalent to measuring complexity in floating point operations (flops).}. To identify an appropriate scale of interest for complexity at high SNR, it is useful to note that in order to achieve a multiplexing gain of $r$ the code must in the high SNR limit have rate of\footnote{Herein, $\log$ denotes the base-2 logarithm and $o(\cdot)$ is the standard Landau notation where $f(\rho) = o(\phi(\rho))$ implies $\lim_{\rho\rightarrow\infty} f(\rho)/\phi(\rho) = 0$. Similarly, $f(\rho) = O(\phi(\rho))$ implies that $\limsup_{\rho\rightarrow\infty} |f(\rho)|/\phi(\rho) < \infty$ for $\phi(\rho) > 0$.} $R = r \log \rho + o(\log \rho)$ bits per channel use where $\rho$ denotes the SNR. Consequently, the cardinality of the codebook is $|\Xset| \doteq \rho^{r T}$ where $T$ is the codeword length and where $\doteq$ denotes equality in the SNR exponent (cf.\ \cite{ZT:03} and Section \ref{sec:notation}). In the worst case, as will be shown later, the sphere decoder is in essence forced to perform a complete search over the entire codebook, and its complexity is in this case $\rho^{rT}$. However, although feasible, this event is also highly improbable. Thus, in order to quantify the probability that the sphere decoder visits a certain (large) number of nodes, we introduce a (complexity) rate-function $\Psi(x)$ over $0 \leq x \leq rT$ given implicitly by $\prob{N \geq \rho^x} \doteq \rho^{-\Psi(x)}$  where $N$ is the complexity of the SD algorithm. In short, $\Psi(x)$ captures the decay-rate of the probability that the complexity exceeds a given SNR dependent threshold $\rho^x$, or equivalently, that the algorithm visits a specific sizable subset of the codebook. This decay-rate should be contrasted with the minimum probability of decoding error, which vanishes in the high SNR limit as $\rho^{-d(r)}$ where $d(r)$ is the diversity gain of the code under maximum likelihood (ML) decoding. We can thus judiciously argue that for any $x$ such that $\Psi(x) > d(r)$, the probability that the complexity exceeds $\rho^x$ is at high SNR insignificant in comparison to the overall probability of error of the decoder. In other words: For $x$ such that $\Psi(x)>d(r)$, imposing a run-time limit of $\rho^x$ on the complexity of the algorithm -- and declaring a decoding outage whenever this limit is not met -- would cause a vanishing degradation in terms of the overall error probability at high SNR. This motivates us to deviate from the traditional worst-case complexity measure that fails to meaningfully describe the effective complexity, and to define the SD complexity exponent $c(r)$ as the infimum of all $x$ for which $\Psi(x) > d(r)$. In essence, $\rho^{c(r)}$ represents the minimum computational reserves required for achieving DMT optimal performance using the SD algorithm. Precise definitions of $c(r)$, and a rigorous treatment of the notion of a vanishing degradation in the overall error probability, is given in Section \ref{sec:decoding-complexity} and by Theorem \ref{thrm:vanishingNew}. The main topic of this work will then be to give closed form expressions, and bounds, for the SD complexity exponent $c(r)$ when decoding different classes of full rate linear codes, to be described later, including the codes proposed in \cite{YW:03,BRV:05,KR:05,EKP:06,ORB:06,ESK:07}.

Most other works on sphere decoding complexity consider uncoded (spatially multiplexed) systems and asymptotic results in terms of the signal space dimension, see, e.g., \cite{JO:04,HV:05a,HV:05b,JO:05}. Our work is instead more related to the analysis in \cite{SJS:09}, which considers the complexity tail distribution for a fixed signal space dimension. However, unlike \cite{SJS:09} we also incorporate the space-time codes into the analysis, as well as the SNR scalings of the rates of these codes mandated by the DMT. In parallel with our work the work in \cite{AD:11} provides an analysis of the complexity tail distribution for unconstrained lattice sequential decoders, in the presence of DMT optimal random lattice codes. A fundamental difference with our work and \cite{AD:11} is that \cite{AD:11} considers unconstrained lattice decoding, whereas we explicitly take into account the constellation boundary in the decoder.  Another difference is that the lattice codes considered in our work can be explicitly constructed and can have arbitrary DMT performance, unlike the random codes in \cite{AD:11} which are non-explicit and which are restricted to being DMT optimal.  We also take our analysis one step further by coupling the complexity tail distribution to the DMT performance of the code in order to obtain the SD complexity exponent. Regarding the ultimate complexity limits on DMT optimal decoding, we have previously established that lattice reduction (LR)-aided linear decoders are sufficient for achieving the entire DMT tradeoff at a worst-case complexity of $O(\log(\rho))$, i.e., corresponding to a complexity exponent of $c(r)=0$ \cite{JE:10}. This is lower than the SD complexity exponent that we will present in what follows. However one notable difference is that the statements made herein are fundamentally stronger in terms of error probability as they not only imply full diversity but also a vanishing SNR gap to the ML decoder (see Theorem \ref{thrm:vanishingNew} for details). Such a result was not established for the decoders in \cite{JE:10,AD:11}.

\subsection{Outline and contributions}

The general definition of the SD complexity exponent is given in Definition~\ref{def:complexityexponent} and Theorem~\ref{thrm:vanishingNew} then describes how sphere decoding and the time-out policies to be employed can guarantee a gap to ML that vanishes with increasing SNR. However, before proceeding with the statement of these results, we first consider the code-channel system, describe the basic workings of the SD algorithm, and handle different pertinent aspects that are necessary for the exposition that follows.

\begin{figure}
\begin{center}
\psfrag{r}[c]{Multiplexing gain $r$}
\psfrag{ce}[c]{Complexity exponent $c(r)$}
\psfrag{n2}[c]{$n=2$}
\psfrag{n3}[c]{$n=3$}
\psfrag{n4}[c]{$n=4$}
\psfrag{n5}[c]{$n=5$}
\psfrag{n6}[c]{$n=6$}
\psfrag{0}{\small 0}
\psfrag{1}{\small 1}
\psfrag{2}{\small 2}
\psfrag{3}{\small 3}
\psfrag{4}{\small 4}
\psfrag{5}{\small 5}
\psfrag{6}{\small 6}
\psfrag{7}{\small 7}
\psfrag{8}{\small 8}
\psfrag{9}{\small 9}
\psfrag{10}{\small 10}
\includegraphics[width=8.3cm]{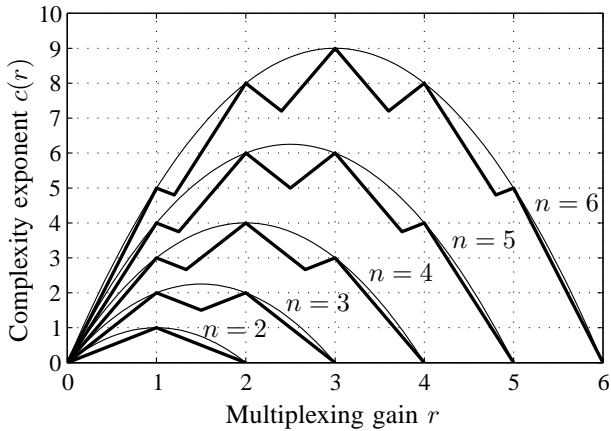}
\end{center}
\caption{The SD complexity exponent $c(r)$ for decoding threaded minimum delay DMT optimal codes with $\nt=T=n$ for $n = 2,\ldots,6$. The SD complexity exponent is illustrated by the bold lines. The same exponent also serves as an upper bound to the SD complexity exponent when decoding any minimum delay DMT optimal full rate linear dispersive code. The thin lines show the quadratic function given by $r(n-r)$ which provides the exact complexity exponent at integer multiplexing gains.
\label{fig:exponents}}
\end{figure}

Following the definition of the SD complexity exponent, Theorem~\ref{thrm:upper-bound} gives, in the form of an optimization problem, a general upper bound $\bar{c}(r)$ on the SD complexity exponent $c(r)$ when decoding any full rate code with multiplexing gain $r$ and diversity $d(r)$. An explicit closed form expression for $\bar{c}(r)$ is then given in Theorem~\ref{thrm:upper-boundDMTopt} for all DMT optimal full rate codes. The bound $\bar{c}(r)$ is already useful in itself in that it establishes that the SD complexity exponent is much lower than the worst-case SNR exponent $rT$ associated with a full search of the codebook. However, in the interest of also establishing the tightness of this bound, Lemma~\ref{lm:lower-bound} provides easy-to-check sufficient conditions on the generator matrix of the code lattice, that guarantee the tightness of $\bar{c}(r)$ in the most general setting. Building on this, Theorem~\ref{thrm:perm-tightness-fullRate} establishes that, given any full rate design of arbitrary DMT performance,  there is always at least one non-random SD column ordering \cite{DGC:03, MGD:06} for which $c(r) = \bar{c}(r)$, i.e., for which the exact $c(r)$ can be explicitly calculated from the result of Theorem~\ref{thrm:upper-bound}. Theorem~\ref{thrm:layered} goes one step further and establishes the exact SD complexity exponent, given any \emph{threaded} code design and the natural column ordering, to be $c(r) = \bar{c}(r)$ while Theorem~\ref{thrm:layeredDMT} provides an explicit expression for $c(r)$ for any DMT optimal threaded code design. Surprisingly, this simple expression (see Fig.~\ref{fig:exponents}) can also serve as an upper bound on $c(r)$ for any full-rate code, irrespective of the fading statistics.  Finally, and along a different path, Theorem~\ref{thrm:nvd} establishes $c(r)$ for any $2\times 2$ approximately universal code \cite{TV:06}, irrespective of its specific structure, thus identifying the exact $c(r)$ even for possibly undiscovered code structural designs, as long as these designs are approximately universal, i.e., as long as they achieve DMT optimality for all fading statistics.  Some general discussions of these results are then provided in Section \ref{sec:discussions}.

The SD complexity exponent for decoding the class of full rate threaded DMT optimal codes with minimum delay, i.e., for which $\nt = T = n$ where $\nt$ denotes the number of transmit antennas, is shown\footnote{A closed form expression for the complexity exponent $c(r)$ shown in Fig.~\ref{fig:exponents} is given by \eqref{eq:layeredDMT} in Theorem~\ref{thrm:layeredDMT}.} in Fig.~\ref{fig:exponents} for $n=2,\ldots,6$. The results shown in the figure apply to codes such as those presented in \cite{YW:03,BRV:05,KR:05,EKP:06,ORB:06,ESK:07}. Before proving the aforementioned results, it is worth commenting on the somewhat counterintuitive result suggested by the SD complexity exponent in Fig.~\ref{fig:exponents}. Namely, that while $c(r)$ initially increases as a function of the multiplexing gain $r$, it then decreases as $r$ approaches its maximum value $\nt$. The initial increase can easily be explained by the fact that the cardinality (and density) of the codebook $\Xset$ increases as a function of $r$. However, the decrease at high multiplexing gains can be understood in light of the coupling of the complexity with the overall probability of error: In short, at high multiplexing gains the error probability is also higher and this implies that the decoder may time out for a larger set of problem instances without seriously affecting the overall performance, leading to an overall reduction in decoding complexity. Pushing the code-decoder pair towards the maximal data rate does therefore not imply that the decoding complexity is maximized. This effect is discussed further in Section~\ref{sec:outages}, in terms of information theoretic outages.

\subsection{Notation} \label{sec:notation}

We let $\ints$, $\reals$, and $\complex$, denote the set of integer, real and complex numbers respectively and $\mathbb{F}^n$ and $\mathbb{F}^{m \times n}$ the set of $n$-vectors and $m \times n$-matrices over $\mathbb{F} \in \{ \ints, \reals, \complex \}$. Vectors are denoted by lower-case bold letters $\bfa$, and matrices are denoted by upper-case bold letters $\bfA$. We use $(\cdot)\tr$ and $(\cdot)\hr$ to denote the transpose and Hermitian (conjugate) transpose of vectors and matrices, and $\vec(\cdot) : \complex^{m \times n} \mapsto \complex^{mn}$ to denote the matrix to vector operation whereby the columns of the argument are stacked on top of each other. We use $\bfI_n \in \complex^{n}$ to denote the $n \times n$ identity matrix, and use $\bfZero$ to denote the zero vector or matrix where the dimensions are given by the context in which $\bfZero$ is used. Deviating slightly from standard usage, we refer to a tall matrix $\bfU \in \complex^{m \times n}$ where $m \geq n$ as unitary if $\bfU\hr\bfU = \bfI$. For $a \in \complex$ we use $\Re(a), \Im(a)$ to respectively denote the real and imaginary parts of $a$. For $a \in \reals$ we use $\lfloor a \rfloor$ to denote the floor operation defined as the largest integer smaller than or equal to $a$,  $\lceil a \rceil$ to denote the ceil operation defined as the smallest integer larger than or equal to $a$, and we let $(a)^+ \defeq \max(a,0)$.

We let $\sigma_1(\bfA) \leq \ldots \leq \sigma_n(\bfA)$ denote the ordered (structurally) non-zero singular values \cite{HJ:85} of a matrix $\bfA \in \complex^{m \times n}$ where $m \geq n$. We will on occasion use $\sigma_{\max}(\bfA)$ to denote the largest singular value when the dimension of $\bfA$ is not explicit. We make no notational difference between random variables and their realizations. We will use the notation $\Omega \defeq \{ \cdots \}$ to label the stochastic event within the brackets.

Finally, in order to simplify notation we will make use of the $\doteq$ (and $\dotleq$, $\dotgeq$, $\dotle$, $\dotge$) notation for equalities (and inequalities) in the SNR exponent, cf.\ \cite{ZT:03}. Specifically, we write $f(\rho) \dotleq \rho^{a}$ and $g(\rho) \dotgeq \rho^{b}$ to denote
$$
\limsup_{\rho\rightarrow\infty} \frac{\log f(\rho)}{\log \rho} \leq a
\quad \text{and} \quad
\liminf_{\rho\rightarrow\infty} \frac{\log g(\rho)}{\log \rho} \geq b
$$
and $f(\rho) \doteq \rho^{x}$ when $f(\rho) \dotleq \rho^{x}$ and $f(\rho) \dotgeq \rho^{x}$ simultaneously hold. The definition of $\dotle$ and $\dotge$ follows after replacing $\leq$ by $<$ and $\geq$ by $>$ in the above expressions.

\section{Channel model and space-time codes}

We consider the standard block Rayleigh fading $\nt \times \nr$ quasi-static point-to-point MIMO channel model with coherence-time $T$ given by
\begin{equation} \label{eq:complex-channel-model}
\bfY = \bfH \bfX + \bfW
\end{equation}
where $\bfX \in \complex^{\nt \times T}$, where $\bfY \in \complex^{\nr \times T}$, and where $\bfW \in \complex^{\nr \times T}$ denote the transmitted space-time block codeword, the block of received signals, and additive spatially and temporally white Gaussian noise. The channel gains $\bfH \in \complex^{\nr \times \nt}$ are assumed to be i.i.d.\ circularly symmetric complex Gaussian\footnote{This assumption is relaxed in Section~\ref{sec:other-fading} where the extension to arbitrary fading distributions is discussed.} (i.e., Rayleigh fading) and constant over the duration of the transmission (i.e., quasi-static Rayleigh fading). We shall assume throughout that $\nr \geq \nt$. The transmitted codewords $\bfX$ are assumed to be drawn uniformly from a codebook $\Xset$ and we assume that
\begin{equation} \label{eq:power-constraint}
\mathrm{E} \{\| \bfX \|^2_\fro \} = \frac{1}{|\Xset|} \sum_{\bfX \in \Xset} \| \bfX \|^2_\fro = \rho T \, ,
\end{equation}
so that the parameter $\rho$ takes on the interpretation of an average SNR. Note also here that one use of \eqref{eq:complex-channel-model} is viewed as $T$ uses of the wireless channel in the definition of the data-rate.

We shall herein restrict our attention to full rate (complex) linear dispersive codes \cite{HH:02,GCD:04} of the form
\begin{equation} \label{eq:ld-form}
\bfX = \theta \sum_{i=1}^\kappa s_{i} \bfD_i
\end{equation}
where $s_i \in \constme \subset \complex$ are constellation symbols drawn from a finite alphabet $\constme$, where $\{ \bfD_i \}_{i=1}^\kappa$ is a set of linearly independent \emph{dispersion matrices}, and where $\theta$ is a parameter regulating the transmit power. The notion of full rate implies that each codeword $\bfX \in \complex^{\nt \times T}$ carries $\kappa = \nt T$ constellation symbols. We will further make the additional assumption that $\constme$ belongs in the class of QAM-like alphabets of the form
\begin{equation} \label{eq:alphabet}
\constme = \const \defeq \, \{ s \, | \, \Re(s), \Im(s) \in \ints \cap [-\eta,\eta] \, \}
\end{equation}
where $\eta > 0$ is a parameter regulating the size of the constellation\footnote{The assumption of a square constellation is made here for simplicity of exposition and in line with practical encoding schemes.  This assumption though can readily be relaxed without affecting the presented results, as long as the constellation is the same for all $s_i, \ i=1,\cdots,\kappa$.  A detailed exposition of the mathematical machinery that allows for this relaxation can be found in \cite[Section III]{JE:10}}. We will use $\constme_\infty$ to denote the extended (infinite) constellation obtained by letting $\eta = \infty$ in \eqref{eq:alphabet}, and note that $\constme_\infty$ is nothing but the set of Gaussian integers. QAM constellations will in general also include a translation and scaling of the underlying lattice $\constme_\infty$. However, as including such a translation would not affect the results obtained herein, and as the scaling can easily be included in the dispersion matrices $\bfD_i$ or $\theta$, we omit these variations and concentrate on \eqref{eq:alphabet} in the interest of notational simplicity.

The channel model in \eqref{eq:complex-channel-model} may also be equivalently expressed in a vectorized form according to
\begin{equation} \label{eq:channel-model}
\bfy = ( \bfI_T \kron \bfH ) \bfx + \bfw
\end{equation}
where $\bfy = \vec(\bfY)$, where $\bfx = \vec(\bfX)$, where $\bfw = \vec(\bfW)$, and where $\kron$ denotes the Kronecker product \cite{HJ:91}. We shall mainly work with \eqref{eq:channel-model} rather than \eqref{eq:complex-channel-model} directly. In the vectorized form the codewords $\bfx$ are given by
$$
\bfx = \theta \bfG \bfs
$$
for $\bfs \in \const^\kappa$, and where the full rank matrix
\begin{equation} \label{eq:generator}
\bfG = \begin{bmatrix}
\vec(\bfD_1) & \cdots & \vec(\bfD_\kappa)
\end{bmatrix}
\end{equation}
is referred to as the \emph{generator matrix} of the code. The linear dispersive codes form a subset of the lattice codes \cite{GCD:04} as the codewords constitute a subset of the (complex) lattice $\theta\bfG \constme_\infty$.

The parameters $\theta$ and $\eta$ are, as noted, chosen in order to satisfy given transmit power and rate constraints. In particular, in order to ensure a multiplexing gain of
\begin{equation} \label{eq:multiplexing}
r \defeq \lim_{\rho \rightarrow \infty} \frac{1}{T} \frac{\log |\Xset|}{\log \rho} \, ,
\end{equation}
or equivalently a rate of $R = r \log \rho + o(\log \rho)$, it must hold that $\eta \doteq \rho^{\frac{r T}{2\kappa}}$ which by \eqref{eq:power-constraint} and \eqref{eq:alphabet} implies that $\theta^2 \doteq \rho^{1-\frac{r T}{\kappa}}$. The code structure described above includes the codes proposed in \cite{YW:03,BRV:05,KR:05,EKP:06,ESK:07}, as well as the QAM-based codes of \cite{ORB:06}, as special cases. Finally, we will throughout, with slight abuse of terminology but still in line with \cite{YW:03,BRV:05,KR:05,EKP:06,ESK:07,ORB:06}, use the term \emph{code} when referring to the whole \emph{family of codes} that is generated by a single generator matrix $\bfG$ for different multiplexing gains and SNRs, and trust that no confusion should follow by this usage.

\section{Decoding}
\label{sec:decoding}

The coherent ML decoder for \eqref{eq:complex-channel-model} is well known to be
\begin{equation} \label{eq:ml}
\bfXh_\ml = \arg \min_{\bfXh \in \Xset} \| \bfY - \bfH \bfXh \|^2_\fro \, .
\end{equation}
The resulting diversity gain of the code, under ML decoding, is correspondingly given by (cf.\ \cite{ZT:03})
\begin{equation} \label{eq:diversity-gain}
d(r) \defeq - \lim_{\rho \rightarrow \infty} \frac{\log \tprob{\bfXh_\ml \neq \bfX}}{\log \rho} \, ,
\end{equation}
where the notation $d(r)$ accentuates the dependence of the diversity on the multiplexing gain $r$.

One of the main features of the linear dispersive codes, as was noted in the introduction, is that their lattice structure allows for efficient -- optimal and near optimal -- solutions to \eqref{eq:ml} using the sphere decoding algorithm. Using the linearity of the map from $\bfs$ to $\bfx = \vec(\bfX)$ we obtain (cf.\ \eqref{eq:channel-model})
\begin{equation} \label{eq:data-moldel}
\bfy = \bfM \bfs + \bfw
\end{equation}
where the code-channel generator matrix $\bfM$ is given by
\begin{equation} \label{eq:equivalent-channel}
\bfM \defeq \theta (\bfI_T \kron \bfH) \bfG \, \in \, \complex^{\nr T \times \kappa} \, .
\end{equation}
We can thus, instead of solving \eqref{eq:ml} directly, equivalently obtain an estimate of $\bfs$ through
\begin{equation} \label{eq:ml-equivalent}
\bfsh_\ml = \arg \min_{\bfsh \in \const^\kappa} \|\bfy - \bfM \bfsh \|^2 \, ,
\end{equation}
where \eqref{eq:ml-equivalent} is an optimization problem suitable for the sphere decoder, and then easily recover $\bfXh_\ml$ from $\bfsh_\ml$.

\subsection{The Sphere Decoder}
\label{sec:sphere-decoder}

The sphere decoding algorithm solves \eqref{eq:ml-equivalent} by a branch-and-bound search on a regular tree. Detailed descriptions of the algorithm are found in \cite{VB:99} and the semi-tutorial papers \cite{AEV:02,DGC:03,MGD:06}, and most implementation issues will not be repeated herein. However, in order to make our results precise and to introduce notation we need to review some or the key ideas as they apply to \eqref{eq:ml-equivalent}.

To this end, note that by the rotational invariance of the Euclidean norm it follows that \eqref{eq:ml-equivalent} is equivalent to
\begin{equation} \label{eq:sd-metric}
\bfsh_\ml = \arg \min_{\bfsh \in \const^\kappa} \| \bfr - \bfR \bfsh \|^2
\end{equation}
where $\bfQ \bfR = \bfM$ is the thin QR-decomposition of $\bfM$ (i.e., $\bfQ \in \complex^{\nr T \times {\kappa}}$ is unitary and $\bfR \in \complex^{\kappa \times \kappa}$ is upper triangular) and where $\bfr = \bfQ\hr\bfy$. The sphere decoder solves \eqref{eq:sd-metric} by enumerating symbol vectors $\bfsh \in \const^{\kappa}$ within a given sphere of radius $\xi > 0$, i.e., $\bfsh$ that satisfy
\begin{equation}  \label{eq:sd-metric-bound}
\| \bfr - \bfR \bfsh \|^2 \leq \xi^2 \, .
\end{equation}
If \eqref{eq:sd-metric-bound} is satisfied for at least one $\bfsh \in \const^\kappa$, then also the ML solution must satisfy \eqref{eq:sd-metric-bound} as the ML solution yields the minimum metric in \eqref{eq:ml-equivalent}. The set of vectors that satisfy \eqref{eq:sd-metric-bound} is found by recursively considering partial symbol vectors $\bfsh_{k} \in \const^k$ for $k=1,\ldots,\kappa$. Specifically, if $\bfsh_{k}$ is the vector containing the last $k$ components of $\bfsh$, a necessary condition for \eqref{eq:sd-metric-bound} to be satisfied is given by
\begin{equation} \label{eq:sd-metric-partial}
\| \bfr_{k} - \bfR_{k} \bfsh_{k} \|^2 \leq \xi^2 \, ,
\end{equation}
where $\bfr_{k} \in \complex^{k}$ denotes the last $k$ components of $\bfr$, and where $\bfR_{k} \in \complex^{k \times k}$ denotes the $k \times k$ lower right corner of $\bfR$. This follows due to the upper triangularity of $\bfR$. Any set of vectors $\bfs \in \const^\kappa$ with common last $k$ components that fail to satisfy \eqref{eq:sd-metric-partial} may be excluded from the set of ML candidate vectors. Enumerating all partial symbol vectors that satisfy \eqref{eq:sd-metric-partial}, beginning with $k=1$, extending these to $k=2$ and so on, yields a recursive procedure for enumerating all $\bfs \in \const^\kappa$ that satisfy \eqref{eq:sd-metric-bound}.

The enumeration of partial symbol vectors $\bfs_{k}$ is equivalent to the traversal of a regular tree with $\kappa$ layers -- one per symbol $s_k$ where $s_k$ is the $k$th component of $\bfs$ -- and $|\const|$ children per node \cite{MGD:06}. There is a one-to-one correspondence between the nodes at layer $k$ (the layers are enumerated with the root node corresponding to $k=0$) and the partial vectors $\bfs_{k}$. We say that a node is visited by the sphere decoder if and only if the corresponding partial vector $\bfs_{k}$ satisfies \eqref{eq:sd-metric-partial}, i.e., there is a bijection between the visited nodes at layer $k$ and the set
\begin{equation} \label{eq:nodes-per-layer}
\Nset_{k} \, \defeq \, \{ \, \bfsh_{k} \in \const^k \; | \;  \| \bfr_{k} - \bfR_{k} \bfsh_{k} \|^2 \leq \xi^2 \, \} \, .
\end{equation}
Due to this relation we will in what follows not make the distinction between nodes and partial symbol vectors and simply refer to $\bfsh_k$ as nodes at layer $k$ when discussing the search. The total number of visited nodes (in all layers of the tree) is given by
\begin{equation} \label{eq:sdcomplexity}
N = \sum_{k=1}^{\kappa} N_{k}  \, ,
\end{equation}
where $N_{k} \defeq |\Nset_{k}|$ is the number of visited nodes at layer $k$ of the search tree. The total number of visited nodes is commonly taken as a measure of the sphere decoder complexity (see \cite{AEV:02,JO:04,JO:05,MGD:06,SJS:09}) and this will also be done in what follows. Note however that as the total number of flops required for evaluating the bound in \eqref{eq:sd-metric-partial} may be upper and lower bounded by constants that are independent of $\rho$ \cite{HV:05a} our results relating to the SD complexity exponent would not change if we instead considered $N$ to be the number of flops spent by the decoder.

\subsection{The search radius}
\label{sec:search-radius}

The description of the sphere decoder is not complete without specifying how the search radius is selected. In the interest of obtaining the SD complexity exponent, we may argue that any reasonable choice of a fixed (non-random) search radius should satisfy
\begin{equation} \label{eq:sphere-radius}
\xi \doteq \rho^0 \, .
\end{equation}
To see this, it is sufficient to note that the metric in \eqref{eq:sd-metric} satisfies
$$
\|\bfr - \bfR \bfs \|^2 = \| \bfQ\hr\bfw \|^2
$$
for the transmitted vector $\bfs$. Thus, if $\|\bfQ\hr\bfw\|^2 > \xi^2$ the transmitted symbol vector is excluded from the search, resulting in a decoding error. By considering a radius that grows slowly with SNR, say $\xi^2 = z \log \rho \doteq \rho^0$, it can be shown that
\begin{equation}\label{eq:sufficientlyLargeRadius}
\tprob{ \|\bfQ\hr\bfw\|^2 \geq \xi^2 } \doteq \rho^{-z} \, ,
\end{equation}
for $z > 0$, i.e., by selecting $z > d(r)$ the probability of excluding the transmitted vector will (for increasing $\rho$) vanish faster than the probability of error and cause vanishing degradation of the overall probability of error. At the same time, if the radius does not tend to infinity with increasing $\rho$, it will follow that $\tprob{ \|\bfQ\hr\bfw\|^2 > \xi^2 }$ is bounded away from zero. This implies a non-vanishing probability of error and a resulting diversity gain of zero, which is clearly undesirable. Thus, as the complexity exponent is not affected by the particular choice of $z$, we shall unless otherwise stated in the following for simplicity assume that $\xi^2 = z \log \rho \doteq \rho^0$, with $z > d(r)$ in order to ensure vanishing degradation to the overall probability of error. This said, the derived SD complexity exponent would be the same if we considered adaptive radius updates as used in the Schnorr-Euchner (SE) implementation \cite{AEV:02,DGC:03}. This may be shown by following the argument in \cite{JO:05}, and we give a proof of this statement in the present setting in Appendix \ref{app:radius-updates}.

\subsection{Decoding Complexity}
\label{sec:decoding-complexity}

The sphere decoder complexity, or equivalently the number of visited nodes $N$, is as stated a random variable with a distribution that depends on a number of parameters, e.g., the system dimensions $\nr$, $\nt$ and $T$, the SNR $\rho$, the multiplexing gain $r$, the generator matrix $\bfG$, and the search radius $\xi$. This is well known and follows by the randomness of the bound in \eqref{eq:sd-metric-partial}. Naturally this randomness must be considered when properly analyzing the sphere decoder complexity, unless one resorts to a worst-case analysis. However, we argue that the worst-case analysis is unnecessarily pessimistic.

In order to illustrate one of the key problems with focusing on the worst-case complexity consider the event that $\bfH = \bfZero$ and $\| \bfw \|^2 < \xi^2$. In this case it is easily seen that \eqref{eq:sd-metric-bound} and \eqref{eq:sd-metric-partial} are always satisfied. As a consequence, the complexity of the sphere decoder would be equal to
$$
N = \sum_{k = 1}^{\kappa} |\const|^k \doteq \sum_{k = 1}^{\kappa} \rho^{\frac{rT}{\kappa}} \doteq \rho^{rT}\, ,
$$
where we have used that $\eta \doteq \rho^{\frac{rT}{2\kappa}}$ to obtain the size of $\const$ in \eqref{eq:alphabet}. The worst-case complexity is therefore comparable to that of a full search over $\Xset$ as $|\Xset|\doteq \rho^{rT}$. However, there is also no point in decoding when $\bfH = \bfZero$ as all codewords would yield the same ML metric which in turn implies a high probability of error. Essentially the same argument, for opting out of decoding, can be made whenever the MIMO channel is in information theoretic outage \cite{ZT:03}. In this case it follows by Fano's inequality that the probability of decoding error will be bounded away from zero. In fact, for a code with a diversity gain of $d(r)$ any set of channel matrices $\Hset$ for which $\prob{\bfH \in \Hset} \dotle \rho^{-d(r)}$, may be neglected by the decoder with vanishing degradation of the overall probability of error. However, rather than identifying and excluding a set of bad channel matrices directly, a more pragmatic approach is to impose a run-time constraint on the decoder and ensure that this constraint is such that the probability of it being violated is insignificant in relation to the probability of error. This leads to the following measure of the decoding complexity, which we will use throughout.

\vspace{3pt}
\begin{definition} \label{def:complexityexponent}
Let
\begin{equation} \label{eq:probability-scaling}
\Psi(x) \defeq - \lim_{\rho \rightarrow \infty} \frac{\log \prob{N \geq \rho^{x}}}{\log \rho} \,
\end{equation}
where $N$ is the number of nodes visited by the sphere decoder (cf.\ \eqref{eq:sdcomplexity}). The \emph{SD complexity exponent} is then given by
\begin{equation} \label{eq:complexity-exponent}
c(r) \; \defeq \; \inf \{ x \, | \,  \Psi(x) > d(r) \}
\end{equation}
where $d(r)$ (cf. \eqref{eq:diversity-gain}) is the diversity gain of the code at multiplexing gain $r$.
\end{definition}
\vspace{3pt}

\subsection{A vanishing gap to the ML performance}
\label{sec:gaptoML}

In order to illustrate the operational significance of $c(r)$, we recall that in addition to the instances where the ML decoder makes an incorrect decision, a time-limited sphere decoder can additionally make decoding errors when the search radius is selected too small, i.e., when $\Nset_{\kappa}=\emptyset$ (cf.\ \eqref{eq:nodes-per-layer}), or when the run-time limit of $\rho^x$ becomes active, i.e., when $N\geq \rho^x$.  These extra errors cause a gap to ML performance which can be quantified as
 \begin{align*}
g(x) & \defeq
\frac{ \tprob{ \{\bfXh_\ml \neq \bfX\} \cup \{ \Nset_{\kappa}=\emptyset \} \cup \{ N \geq \rho^x \}} }
{ \tprob{\bfXh_\ml \neq \bfX} }
\end{align*} describing the ratio between the probability of error of the time-limited sphere decoder and the ML decoder. With respect to $c(r)$ we then have the following.
\vspace{3pt}
\begin{theorem} \label{thrm:vanishingNew}
A sphere decoder with a computational constraint activated at $\rho^x$ flops, allows for a vanishing gap to ML performance for all $x>c(r)$, i.e., \begin{equation} \label{eq:gap1}
\lim_{\rho \rightarrow \infty} g(x) = 1,  \ \text{for any} \ x > c(r).
\end{equation}
\end{theorem}
\vspace{3pt}
The above simply states that for any $x > c(r)$ it is possible to design a decoder based on the SD algorithm that achieves a vanishing SNR gap to the ML decoder, at a worst-case complexity of $\rho^x$.
To see this apply the union bound to get
\begin{align*}
g(x) & \leq \underbrace{\frac{ \tprob{ \bfXh_\ml \neq \bfX } }
{ \tprob{\bfXh_\ml \neq \bfX} }}_{=1} +
\underbrace{\frac{ \tprob{ \xi_\ml > \xi} }
{ \tprob{\bfXh_\ml \neq \bfX} }}_{\rightarrow 0} +
\underbrace{\frac{ \tprob{ N \geq \rho^x} }
{ \tprob{\bfXh_\ml \neq \bfX} }}_{\rightarrow 0}
\end{align*}
where the second and third term tend to zero with increasing $\rho$ as the numerator tends to zero at a faster rate than the denominator cf. \eqref{eq:sufficientlyLargeRadius}, \eqref{eq:complexity-exponent}. This immediately translates to a vanishing SNR gap to the ML decoder at high SNR. In short, the probabilities of the events that the search space is empty or that the complexity of the run-time-unconstrained sphere decoder exceeds $\rho^x$ are insignificant in comparison to the probability of ML decoding errors.

Furthermore,  one cannot in general time-limit the sphere decoder to $\rho^x$ for some $x < c(r)$ and expect an arbitrary small gap to ML performance.  Specifically, one can show (cf.\ \eqref{eq:lower-bound-2}) that $\tprob{ N \geq \rho^x } \dotge \rho^{-d(r)}$ for any $x < c(r)$, and as a result, it follows\footnote{Note here that what we formally show is that under the basic technical conditions of Lemma~\ref{lm:lower-bound} one cannot time-limit the decoder to $\rho^x$ for any $x < c(r)$. The same statement naturally holds whenever $\Psi(x)$ is strictly decreasing in $x$.} for $x < c(r)$ that
$$
g(x)\geq \underbrace{\frac{ \tprob{ N \geq \rho^x} }
{ \tprob{\bfXh_\ml \neq \bfX} }}_{\rightarrow \infty}
$$
implying that any attempt to significantly reduce the complexity below $\rho^{c(r)}$ will be at the expense of the vanishing SNR gap to ML decoding.

\section{The sphere decoder complexity exponent} \label{sec:upper-bound}

We proceed to establish upper and lower bounds on the SD complexity exponent, in essence through the application of a principle (dating back to Gauss) which states that the number of integer lattice points within a (large) set is well approximated by the volume of the set \cite{Kra:88,GW:93}. Thus, in order to approximate the number of nodes at layer $k$ of the search tree, i.e., the size of $\Nset_k$ defined in \eqref{eq:nodes-per-layer}, we are primarily concerned with the volume of $([-\eta,\eta]+\sqrt{-1}[-\eta,\eta])^k \cap \Eset_k$ where $\Eset_k$ is the elliptical set given by (cf. \eqref{eq:nodes-per-layer})
\begin{equation} \label{eq:partial-ellipsoid}
\Eset_k = \{ \bfst_k \in \complex^k \, | \, \|\bfr_k - \bfR_k\bfst_k \|^2 \leq \xi^2 \, \} \, .
\end{equation}
The use of the volume principle for assessing the sphere decoder complexity was previously used in \cite{BK:98,AEV:02,SJS:09} although its prior use in the communications literature is limited to the case of lattice decoding (i.e., where the constellation boundary constraint imposed by $[-\eta,\eta]$ is ignored by the decoder). Herein, we have to take the constellation boundary into account to obtain tight bounds on the SD complexity exponent.

The upper and lower bounds presented in this section are essentially obtained in three main steps: 1) The volume principle is used to obtain an expression for the number $N_k$ of visited nodes at layer $k$ in terms of the singular values of $\bfR_k$; 2) the singular values of $\bfR_k$ for $k=1,\ldots,\kappa$ are related to the singular values of the channel matrix $\bfH$; and 3) the theory of large deviations is used similarly to \cite{ZT:03} to identify random events likely to cause an atypically large decoding complexity. Establishing the upper bound on $c(r)$ turns out to be easier mathematically. The reason for this is primarily in the second step where the interlacing property of singular values of sub-matrices \cite{HJ:85} can be used to lower bound the singular values of $\bfR_k$ by the singular values of $\bfH$, to yield results that are universally applicable for any full rank generator matrix $\bfG$, cf.\ Theorem \ref{thrm:upper-bound}. Although the interlacing property gives both upper and lower bounds on the singular values of $\bfR_k$, the upper bounds are unfortunately not sufficient for establishing tight lower bounds on $c(r)$. We are therefore forced to develop tighter bounds that depend on some technical assumptions on $\bfG$, cf.\ Lemma~\ref{lm:lower-bound}. While these conditions are, at least in principle, easily verified for any given code design, they are generally hard to verify for arbitrary classes of codes. Nevertheless, for some important classes discussed in Section \ref{sec:tightness} we are able to conclude that the upper bound on $c(r)$ is tight, thereby establishing $c(r)$ exactly.

The derivation of the upper bound on $c(r)$ is given in the following sub-sections, while the derivation of the lower bound, which is similar in spirit to the upper bound but complicated by some technical details, is primarily given in Appendix \ref{app:lower-bound}, and discussed in Section \ref{sec:tightness}.

\subsection{The volume principle}

As noted, we begin by establishing bounds on $N_k = |\Nset_k|$ in terms of the singular values of the matrix $\bfR_k$ in \eqref{eq:sd-metric-partial}, the sphere radius $\xi$ and the constellation size $\eta$. To this end, consider the following lemma, which corresponds to rigorous applications of the volume principle discussed above. The proof of the lemma is given in Appendix \ref{app:upper-bound}.

\vspace{3pt}
\begin{lemma} \label{lm:upper-bound}
Let $\Eset \subset \reals^n$ be the ellipsoidal set given by
\begin{equation} \label{eq:lm-elipsoid}
\Eset \defeq \{ \bfd \in \reals^n \; | \; \| \bfc - \bfD \bfd \|^2 \leq \xi^2 \}
\end{equation}
where $\bfD \in \reals^{n \times n}$ and $\bfc \in \reals^n$. Let $\Bset \subset \reals^n$ be the hypercube given by
\begin{equation} \label{eq:lm-hypercube}
\Bset \defeq \{ \bfd \in \reals^n \; | \; |d_i| \leq \eta \, , \, i=1,\ldots,n \, \} \, .
\end{equation}
Then, the number of integer points contained in the intersection of $\Eset$ and $\Bset$ is upper bounded as
\begin{equation} \label{eq:lm-upper}
| \Eset \cap \Bset \cap \ints^n | \leq \prod_{i=1}^n \left[ \sqrt{n} + \min\left( \frac{2 \xi}{\sigma_{i}(\bfD)} \, , \, 2 \sqrt{n} \eta \right )  \right] \, ,
\end{equation}
and the number of integer points contained in $\Eset$ is lower bounded by
\begin{equation} \label{eq:lm-lower}
| \Eset \cap \ints^n | \geq \prod_{i=1}^n \left( \frac{2 \xi}{\sqrt{n} \sigma_{i}(\bfD)} - \sqrt{n} \right )^{+\!} \, ,
\end{equation}
where $\sigma_{i}(\bfD)$, $i=1,\ldots,n$ denote the singular values of $\bfD$.
\end{lemma}
\vspace{3pt}

Although Lemma \ref{lm:upper-bound} is phrased in terms of real valued quantities, it is easily applied to complex valued sets by considering each complex dimension as two real valued dimensions. In particular, the expression in \eqref{eq:sd-metric-bound} is equivalent to
$$
\| \underline{\bfr}_k - \underline{\bfR}_k \underline{\bfsh}_k \|^2 \leq \xi^2
$$
where
$$
\underline{\bfr}_k = \begin{bmatrix}
\Re(\bfr_k) \\ \Im(\bfr_k)
\end{bmatrix} , \,
\underline{\bfR}_k = \begin{bmatrix}
\Re(\bfR_k) & \hspace{-5pt} -\Im(\bfR_k) \\
\Im(\bfR_k) & \Re(\bfR_k) \\
\end{bmatrix} ,
$$
and
$$
\underline{\bfsh}_k = \begin{bmatrix}
\Re(\bfsh_k) \\ \Im(\bfsh_k)
\end{bmatrix} \, .
$$
By noting that if $\bfR_k = \bfU\bfSigma\bfV\hr$ is the singular value decomposition (SVD) \cite{HJ:85} of $\bfR_k$, then
$$
\underline{\bfR}_k =
\begin{bmatrix}
\Re(\bfU) & \hspace{-5pt} -\Im(\bfU) \\
\Im(\bfU) & \Re(\bfU) \\
\end{bmatrix}
\begin{bmatrix}
\bfSigma & \bfZero \\
\bfZero & \bfSigma \\
\end{bmatrix}
\begin{bmatrix}
\Re(\bfV\hr) & \hspace{-5pt} -\Im(\bfV\hr) \\
\Im(\bfV\hr) & \Re(\bfV\hr) \\
\end{bmatrix}
$$
is an SVD of $\underline{\bfR}_k$, it follows that the singular values of $\underline{\bfR}_k$ are the same as those of $\bfR_k$ albeit with a multiplicity of $2$. Thus, applying \eqref{eq:lm-upper} in Lemma~\ref{lm:upper-bound} to $\Nset_{k}$ (cf.\ \eqref{eq:nodes-per-layer}) yields an upper bound on the number of nodes visited at layer $k$, which is given as
\begin{equation} \label{eq:nodes-per-layer-ub}
N_k = | \Nset_{k} | \leq \prod_{i=1}^{k} \left[ \sqrt{2k} + \min \left( \frac{2 \xi }{\sigma_{i}(\bfR_{k})} \, , 2 \sqrt{2k} \eta \right) \right]^2
\end{equation}
where $\sigma_{i}(\bfR_k)$, $i=1,\ldots,k$ denote the singular values of $\bfR_k$. Here, in essence, the additive $\sqrt{2k}$ term accounts for edge effects in the volume approximation, the first term in the minimum accounts for the size of the search sphere, and the second term in the minimum accounts for the constellation boundary.

The lower bound in \eqref{eq:lm-lower} will be used later in order to assess the tightness of the upper bound on $c(r)$ developed next. The reason for providing a lower bound on $| \Eset \cap \ints^n |$ and not $| \Eset \cap \Bset \cap \ints^n |$ is that we cannot a-priori rule out that $\bfc$ in \eqref{eq:lm-elipsoid} is such that $ \Eset \cap \Bset = \emptyset$, a case which if not ruled out would lead to the trivial lower bound $| \Eset \cap \Bset \cap \ints^n | \geq 0$.

\subsection{Singular value bounds}

The interlacing theorem of singular values of sub-matrices (cf.\ \cite[Th.\ 7.3.9]{HJ:85} and\cite[Corollary 3.1.3]{HJ:91}) states that the singular values of $\bfR_k$ are bounded by the singular values of $\bfR$ according
\begin{equation} \label{eq:interlace}
\sigma_{i+\kappa-k}(\bfR) \geq \sigma_i(\bfR_k) \geq \sigma_i(\bfR) \, , \quad i=1,\ldots,k \, ,
\end{equation}
where $\sigma_i(\bfR_k)$ and $\sigma_i(\bfR)$ denotes the $i$th singular values of $\bfR_k$ and $\bfR$ respectively. As $\bfR = \bfQ\hr\bfM$ where $\bfQ$ has a set of orthogonal columns that span the range of $\bfM$ it follows that $\sigma_i(\bfR) = \sigma_i(\bfM)$. Further, by the definition of $\bfM$ in \eqref{eq:equivalent-channel} we have that $\sigma_i(\bfM) \geq \theta \gamma \sigma_i(\bfI_T \kron \bfH)$ where $\gamma \defeq \sigma_1(\bfG) > 0$ due to the assumption that $\bfG$ is full rank. The singular values of $\bfI_T \kron \bfH$ are the same as those of the channel matrix $\bfH$ in \eqref{eq:complex-channel-model}, albeit with a multiplicity of $T$, i.e.,
$$
\sigma_i(\bfI_T \kron \bfH) = \sigma_{\iota_{T}(i)}(\bfH) \, , \quad i=1,\ldots,\nt \, ,
$$
where
\begin{equation} \label{eq:multiplicity}
\iota_T(i) \defeq \left\lceil \frac{i}{T} \right\rceil \, .
\end{equation}
This can be seen by noting that if $\bfH = \bfU\bfSigma\bfV\hr$ is the SVD of $\bfH$, then
$$
(\bfI_T \kron \bfU)(\bfI_T \kron \bfSigma)(\bfI_T \kron \bfV\hr)
$$
is an SVD of $\bfI_T \kron \bfH$ (albeit with a non-standard ordering of the singular values). Alternatively, on can apply \cite[Theorem 4.2.12]{HJ:91} to the eigenvalues of $\bfI_T \kron \bfH\hr\bfH$.

Combining the above yields a lower bound on the singular values of $\bfR_k$ in terms of the singular values of the channel matrix $\bfH$ according to
$$
\sigma_i(\bfR_k) \geq \theta \gamma \sigma_{\iota_{T}(i)}(\bfH) \, , \quad i=1,\ldots,k \, ,
$$
and an upper bound on the number of nodes visited by the sphere decoder at layer $k$ according to
\begin{equation} \label{eq:nodes-per-layer-ub2}
N_k \leq \prod_{i=1}^{k} \left[ \sqrt{2k} + \min \! \left( \frac{2 \xi }{\theta \gamma \sigma_{\iota_{T}(i)}(\bfH)} \, , 2 \sqrt{2k} \eta \right) \right]^2 .
\end{equation}
In order to bound the probability that the right hand side of \eqref{eq:nodes-per-layer-ub2} is atypically large in the high SNR regime, it is useful to consider the SNR dependent parameterization of the singular values (or eigenvalues) of $\bfH\hr\bfH$ introduced in \cite{ZT:03}, i.e., SNR dependent random variables $\alpha_{i}$, for $i=1,\ldots,\nt$, defined by
\begin{equation} \label{eq:eigenvalue-asymptot}
\alpha_{i} \defeq -\frac{\log \sigma_{i}(\bfH\hr\bfH)}{\log \rho}
\quad \Leftrightarrow \quad
\sigma_{i}(\bfH\hr\bfH) = \rho^{-\alpha_{i}}  \, .
\end{equation}
Note that by this definition $\sigma_i(\bfH) = \rho^{-\frac{1}{2}\alpha_i}$. The variables, $\alpha_i$ for $i=1,\ldots,\nt$ are referred to as the \emph{singularity levels} of $\bfH$ as they give an indication of how close to singular the channel $\bfH$ is in relation to the inverse SNR $\rho^{-1}$. As $\xi \doteq \rho^0$, $\theta \doteq \rho^{\frac{1}{2}-\frac{rT}{2 \kappa}}$ and $\eta = \rho^{\frac{r T}{2 \kappa}}$ it holds that
$$
\left[
\sqrt{k} + \min \! \left( \frac{2 \xi }{\theta \gamma \sigma_{\iota_{T}(i)}(\bfH)} \, , 2 \sqrt{k} \eta \right) \right]^2 \dotleq \rho^{\nu_{i}}
$$
where
\begin{equation} \label{eq:nui-def}
\nu_{i} \defeq \min\!\left( \frac{rT}{\kappa} - 1 + \alpha_{\iota_{T}(i)} \, , \, \frac{rT}{\kappa} \right)^{\!+}.
\end{equation}
By \eqref{eq:nodes-per-layer-ub2} it follows that
\begin{equation} \label{eq:layer-node-asbound}
N_{k} \dotleq \prod_{i=1}^{k} \rho^{\nu_{i}} = \rho^{\sum_{i=1}^{k} \nu_{i}} \, .
\end{equation}
However, as the SNR exponent on the right hand side of \eqref{eq:layer-node-asbound} is non decreasing in $k$ it must for the total number of visited nodes $N$ hold that $N = \sum_{k=1}^{\kappa} N_{k} \dotleq \rho^{\sum_{i=1}^{\kappa} \nu_{i}}$ or for any given $\delta > 0$ hold that
\begin{equation} \label{eq:total-nodes-ub}
N \leq \rho^{\sum_{i=1}^{\kappa} \nu_{i} + \delta}
\end{equation}
provided $\rho$ is sufficiently large.

Consider now the set (cf.\ \eqref{eq:nui-def} and \eqref{eq:total-nodes-ub})
\begin{equation} \label{eq:decoding-outage-set}
\Tset(x) \defeq \left\{ \bfalpha \; \Big| \;
\sum_{i=1}^{\kappa} \min\!\left( \frac{rT}{\kappa} - 1 + \alpha_{\iota_{T}(i)} \, , \, \frac{rT}{\kappa} \right)^{\!+} \geq x
 \right\} \, ,
\end{equation}
where $\bfalpha = (\alpha_{1},\ldots,\alpha_{\nt})$.
As \eqref{eq:total-nodes-ub} holds (asymptotically) for any $\delta > 0$, and since $\bfalpha \notin \Tset(y)$ implies that $N < \rho^x$ for any $y < x$ by \eqref{eq:total-nodes-ub} and \eqref{eq:nui-def}, it follows that
$$
\lim_{\rho \rightarrow \infty} \frac{\log \prob{N \geq \rho^{x}}}{\log \rho} \leq \lim_{\rho \rightarrow \infty} \frac{\log \prob{\bfalpha \in \Tset(y)}}{\log \rho} \, .
$$
Equivalently (cf.\ \eqref{eq:probability-scaling})
\begin{equation} \label{eq:scaling-lower-bound}
\Psi(x) \geq - \lim_{\rho \rightarrow \infty} \frac{\log \prob{\bfalpha \in \Tset(y)}}{\log \rho}
\end{equation}
for $y < x$. The value of the bound in \eqref{eq:scaling-lower-bound} is that the right hand side is readily computed using large deviation theory  \cite{DZ:98}.

\subsection{Large deviations}

A sequence of random vectors $\bfbeta_{\rho} \in \reals^n$ parameterized by $\rho$ is said to satisfy the \emph{large deviation principle} \cite{DZ:98} with \emph{rate function} $I$,
$$
I : \reals^n \mapsto \{ \reals_{+} \, , \, \infty \} \, ,
$$
if for every open set $\Gset \subseteq \reals^n$ it holds that
\begin{equation} \label{eq:ld-open-bound}
\liminf_{\rho \rightarrow \infty} \frac{\log \prob{\bfbeta_{\rho} \in \Gset}}{\log \rho} \geq - \inf_{\bfbeta \in \Gset} I(\bfbeta)
\end{equation}
and if for every closed set $\Fset \subseteq \reals^n$ it holds that
\begin{equation} \label{eq:ld-closed-bound}
\liminf_{\rho \rightarrow \infty} \frac{\log \prob{\bfbeta_{\rho} \in \Fset}}{\log \rho} \leq - \inf_{\bfbeta \in \Fset} I(\bfbeta) \, .
\end{equation}
Although not stated formally, one of the central results of \cite{ZT:03} is that the sequence of random variables given by $\bfalpha_{\rho} = \bfalpha = (\alpha_{1},\ldots,\alpha_{\nt})$ (cf. \eqref{eq:eigenvalue-asymptot}) satisfies the large deviation principle with rate function (see the proof of Theorem 4 in \cite{ZT:03})
\begin{equation} \label{eq:rate-function}
I(\bfalpha) = \sum_{i=1}^{\nt} (\nr-\nt+2i-1)\alpha_{i}
\end{equation}
if $\alpha_{1} \geq \ldots \geq \alpha_{\nt} \geq 0$ and $I(\bfalpha) = \infty$ otherwise. This observation was key in establishing the DMT in \cite{ZT:03}.

By combining \eqref{eq:scaling-lower-bound} with \eqref{eq:ld-closed-bound}, and noting that $\Tset(y)$ is a closed set, it follows that
\begin{equation} \label{eq:fy-def}
\Psi(x) \geq f(y) \defeq \inf_{\bfalpha \in \Tset(y)} I(\bfalpha)
\end{equation}
for any $y < x$. As $\Tset(x) \subseteq \Tset(y)$ for all $y \leq x$ it follows that $f(y)$ is non-decreasing and it can additionally be verified that $f(y)$ is left-continuous, i.e.,
$$
\sup_{y < x} f(y) = f(x) \, ,
$$
which implies that
\begin{equation} \label{eq:final-gamma-lower-bound}
\Psi(x) \geq f(x) = \inf_{\bfalpha \in \Tset(x)} I(\bfalpha) \, .
\end{equation}
From \eqref{eq:final-gamma-lower-bound} it follows that the complexity exponent $c(r)$ is upper bounded by $\bar{c}(r)$ where
\begin{equation} \label{eq:opt0}
\bar{c}(r) \defeq \inf \{ x \, | \, f(x) > d(r) \} = \sup \{ x \, | \, f(x) \leq d(r) \}
\end{equation}
and where the last equality follows as $f(x)$ is non-decreasing. Further, by the left-continuity of $f(x)$ it follows that the supremum on the right is attained, i.e., the supremum can be replaced by a maximum.

Note however that the condition that $f(x) \leq d(r)$ is satisfied if and only if there exist an $\bfalpha \in \Tset(x)$ such that $I(\bfalpha) \leq d(r)$. Thus, $\bar{c}(r)$ in \eqref{eq:opt0} could alternatively be obtained as the solution to a constrained maximization problem according to
\begin{subequations} \label{eq:opt1}
\begin{align}
\max_{\bfalpha, x} \quad & x \\
\text{s.t.} \quad & \sum_{i=1}^{\kappa} \min\! \left( \frac{rT}{\kappa} - 1 +  \alpha_{\iota_{T}(i)} \, , \, \frac{rT}{\kappa} \right)^{\!+} \geq x \label{eq:opt1-set}\\
& \sum_{i=1}^{\nt} (\nr-\nt+2i - 1)\alpha_{i} \leq d \label{eq:opt1-div} \\
& \alpha_{1} \geq \ldots \geq \alpha_{\nt} \geq 0 \, , \label{eq:opt1-implicit}
\end{align}
\end{subequations}
where \eqref{eq:opt1-set} follows from the constraint $\bfalpha \in \Tset(x)$, and where \eqref{eq:opt1-div} and \eqref{eq:opt1-implicit} follows from $I(\bfalpha) \leq d(r)$. It is straightforward to show that the optimal $x$ in \eqref{eq:opt1} must be such that \eqref{eq:opt1-set} is satisfied with equality. By further noting that the sum in \eqref{eq:opt1-set} contains only $\nt$ distinct terms, each with multiplicity $T$, it can be seen that
\begin{align*}
& \sum_{i=1}^{\kappa} \min
 \left( \frac{rT}{\kappa} - 1 + \alpha_{\iota_{T}(i)} \, , \, \frac{rT}{\kappa} \right)^{\!+} \\
 = & \sum_{i=1}^{\nt}  T \min\!
 \left( \frac{r}{\nt} - 1 + \alpha_{i}\, , \, \frac{r}{\nt} \right)^{\!+} \, ,
\end{align*}
where we have also used the full rate assumption that $\kappa = \nt T$. We summarize the above in the following theorem.

\vspace{5pt}
\begin{theorem} \label{thrm:upper-bound}
The SD complexity exponent $c(r)$ of decoding any full rate linear dispersive code with multiplexing gain $r$ and diversity $d(r)$ is upper bounded as $c(r) \leq \bar{c}(r)$ where
\begin{subequations}  \label{eq:optimization}
\begin{align}
\bar{c}(r) \, \defeq \, \max_{\bfalpha} \; & \sum_{i=1}^{\nt}  T \min \!\left( \frac{r}{\nt} - 1 + \alpha_{i}\, , \, \frac{r}{\nt} \right)^{\!+}\label{eq:optimization-objective} \\
\text{s.t.} \; &  \sum_{i=1}^{\nt} (\nr-\nt+2i - 1)\alpha_{i} \leq d(r) \label{eq:optimization-div} \\
 & \alpha_{1} \geq \ldots \geq \alpha_{\nt} \geq 0 \, . \label{eq:optimization-pos}
\end{align}
\end{subequations}
\end{theorem}
\vspace{5pt}

The upper bound given by Theorem \ref{thrm:upper-bound} can naturally be computed given explicit values for the multiplexing gain $r$ and diversity gain\footnote{Note that Theorem \ref{thrm:upper-bound} does not assume a diversity optimal code.} $d(r)$. However, it is also possible in some cases to give general solutions as a function of $r$ when the DMT curve $d(r)$ of the code is known explicitly. In particular, DMT optimal codes such as those presented in \cite{YW:03,BRV:05,KR:05,EKP:06,ORB:06,ESK:07} have a diversity gain of $d(k) = (\nt - k)(\nr-k)$ at any integer multiplexing gain $r=k$ \cite{ZT:03}. In this case it is straightforward to verify that an\footnote{In general, \eqref{eq:optimization} does not have a unique optimal point as $\min(a,b)^{+} $ is constant in $a$ for $a \leq 0$ and $a \geq b$.} optimal $\bfalpha$ in \eqref{eq:optimization} is given by
$$
\alpha_{i}^\star = 1, \quad \text{for~} i=1,\ldots,\nt-k
$$
and
$$
\alpha_{i}^\star = 0, \quad \text{for~} i=\nt-k+1,\ldots,\nt \, .
$$
To see this, note that the objective function in \eqref{eq:optimization-objective} is symmetric with respect to permutations of the set of $\alpha_{i}$ for $i=1,\ldots,\nt$. As the sum in the diversity constraint \eqref{eq:optimization-div} places more weigh on $\alpha_{j}$ than on $\alpha_{i}$, for $j > i$, it is optimal to increase $\alpha_{1}$ until the term in \eqref{eq:optimization-objective} containing $\alpha_{1}$ saturates (i.e., when $\alpha_{1} = 1$),  then to increase $\alpha_{2}$ etcetera, until the constraint is satisfied with equality. This yields the aforementioned solution. Note also that $\alpha_1^\star,\ldots,\alpha_\nt^\star$ are the same singularity levels that give the typical outages in \cite{ZT:03}, (cf.\ Section \ref{sec:outages}). Inserting the optimal solution into \eqref{eq:optimization-objective} yields
$$
\bar{c}(k) = \frac{T k (\nt-k) }{\nt} \, ,
$$
which is a remarkably simple upper bound on the SD complexity exponent of decoding any DMT optimal code at an integer multiplexing gain $r=k$.

For a DMT optimal code at a possibly non-integer value of $r$, let $k$ be the integer such that $r \in [k,k+1)$, i.e., $k = \lfloor r \rfloor$. The optimal solution is in this case given by
$$
\alpha_{i}^\star = 1, \quad \text{for~} i=1,\ldots,\nt-k-1 \, ,
$$
$$
\alpha_{i}^\star = 0, \quad \text{for~} i=\nt-k+1,\ldots,\nt \, ,
$$
and
$$
\alpha_{\nt-k}^\star = k+1-r \, .
$$
Substituting the above solution back into \eqref{eq:optimization-objective} yields
$$
\bar{c}(r) = \frac{T}{\nt} \Big( r(\nt-k-1) + \big( \nt k - r(\nt-1) \big)^+ \Big) \, .
$$
We summarize the above in the following theorem.

\vspace{5pt}
\begin{theorem} \label{thrm:upper-boundDMTopt}
The SD complexity exponent $c(r)$ of decoding any DMT optimal full rate linear dispersive code with integer multiplexing gain $r=k$ is upper bounded as
\begin{equation} \label{eq:final-upper-bound-int}
c(k) \leq \bar{c}(k) = \frac{T k (\nt-k) }{\nt} \, .
\end{equation}
For general $r$ where $0 \leq r \leq \nt$ the SD complexity exponent $c(r)$ is upper bounded as $c(r) \leq \bar{c}(r)$ where
\begin{equation} \label{eq:final-upper-boundDMTopt}
\bar{c}(r) = \frac{T}{\nt} \Big( r(\nt-\lfloor r \rfloor-1) + \big( \nt \lfloor r \rfloor - r(\nt-1) \big)^+ \Big) \, .
\end{equation}
\end{theorem}
\vspace{5pt}

The function $\bar{c}(r)$ in \eqref{eq:final-upper-boundDMTopt} is a piecewise linear function in $r$, although slightly more involved than the set of straight lines describing the optimal DMT $d(r)$. For $\nt = T = n$ the function in \eqref{eq:final-upper-boundDMTopt} coincides with the curve for $c(r)$ shown in Fig.~\ref{fig:exponents}.

\subsection{Establishing the exact SD complexity exponent} \label{sec:tightness}

We now turn to specific cases where we can exactly establish the SD complexity exponent $c(r)$ by establishing that the upper bound $c(r) \leq \bar{c}(r)$ is in fact tight. To this end, we begin with the following lemma, which provides a sufficient condition for $c(r) = \bar{c}(r)$, i.e., for the tightness of the upper bound.

\vspace{3pt}
\begin{lemma} \label{lm:lower-bound}
Let $\bfG_{|p} \in \complex^{\kappa \times Tp}$ be the matrix consisting of the first $Tp$ columns of the generator matrix $\bfG \in \complex^{\kappa \times \kappa}$. If there exists, for $p=1,\ldots,\nt$, unitary matrices $\bfU_p \in \complex^{\nt \times p}$ such that
\begin{equation} \label{eq:rank-condition}
\rank\big( (\bfI_T \kron \bfU_p\hr) \bfG_{|p} \big) = pT \, ,
\end{equation}
then $c(r) = \bar{c}(r)$ for all $r \in [0,\nt]$, where $\bar{c}(r)$ is given by \eqref{eq:optimization} in Theorem \ref{thrm:upper-bound}.
\end{lemma}
\vspace{3pt}

The proof of Lemma~\ref{lm:lower-bound} is similar in spirit to the proof of Theorem~\ref{thrm:upper-bound}, although riddled with technical details, and therefore relegated to Appendix~\ref{app:lower-bound}. In essence, the condition posed in \eqref{eq:rank-condition} implies that there are certain orientations of the right singular vectors of the channel $\bfH$ (in relation to the code generated by $\bfG$) for which the lower bound in \eqref{eq:interlace} is sufficiently tight. Details are provided in Appendix \ref{app:lower-bound}, and some additional discussions of \eqref{eq:rank-condition} and the general applicability of the lemma can be found in Section~\ref{sec:discussions}. However, we first apply Lemma \ref{lm:lower-bound} to find $c(r)$ in some very important special cases.

To this end, it is useful to first note that permuting the columns of $\bfG$, i.e., replacing $\bfG$ with $\bfG\bfPi$ where $\bfPi \in \reals^{\kappa \times \kappa}$ is a permutation matrix, does not change the code $\Xset$. Instead, the effect such a permutation would have is that it would change the order in which the symbols in $\bfs$ are enumerated by the sphere decoder described in Section \ref{sec:sphere-decoder} (cf. \cite[Section IV]{DGC:03}). In the present context, the first $pT$ columns of $\bfG\bfPi$, i.e., $[\bfG\bfPi]_{|p}$, may differ from those of $\bfG$. Thus, we see that \eqref{eq:rank-condition} depends not only on the code itself, but also on the order in which the constituent symbols $s_i$ are enumerated by the sphere decoder (cf.\ \cite{DGC:03,MGD:06} where the topic of column ordering is discussed in detail).

In the context of Lemma \ref{lm:lower-bound}, it can be seen that as $(\bfI_T \kron \bfU_p\hr) \bfG \in \complex^{nT \times \kappa}$ has rank $pT$ for any unitary $\bfU_p$ due to the full rank assumption on $\bfG$. One can therefore select $pT$ linearly independent columns, or equivalently, find a permutation matrix $\bfPi$ such that $(\bfI_T \kron \bfU_p\hr) [\bfG\bfPi]_{|p}$ has full rank. Using a similar argument, we can recursively construct a (single) $\bfPi$ for which there are $\bfU_p$ for $p=1,\ldots,\nt$ satisfying
$$
\rank \big( (\bfI_T \kron \bfU_p\hr) [\bfG\bfPi]_{|p} \big) = pT
$$
by constructing $\bfU_{p-1}$ from $\bfU_p$ by removing a column, selecting the appropriate columns from $\bfG$, and starting the recursion with an arbitrary $\bfU_\nt$. Interpreting the above in light of Theorem~\ref{thrm:upper-bound} and Lemma~\ref{lm:lower-bound} we can thus establish that for any full rate linear dispersive code design, $\bar{c}(r)$ as defined in Theorem~\ref{thrm:upper-bound} and given in Theorem~\ref{thrm:upper-boundDMTopt} for DMT optimal codes, is the tightest upper bound on the SD complexity exponent that can possibly hold under arbitrary column orderings. This is formalized in the following.

\vspace{3pt}
\begin{theorem} \label{thrm:perm-tightness-fullRate}
Given any full rate linear dispersive code achieving diversity $d(r)$, there is always at least one column ordering for which $c(r) = \bar{c}(r)$.
\end{theorem}
\vspace{3pt}

However, while Theorem~\ref{thrm:perm-tightness-fullRate} is useful in the sense that it tells us that one could not improve upon the tightness of $\bar{c}(r)$ without introducing further assumptions regarding the particular code design considered, it is obviously not of practical interest to use the worst possible column ordering. Therefore, we turn our attention to the important class of threaded codes \cite{GD:03} for which we will show that the \emph{natural} column ordering $\bfPi = \bfI_\kappa$ implies $c(r) = \bar{c}(r)$.

\subsubsection{Threaded codes} \label{sec:threaded}

The class of threaded code designs is of particular interest, as it includes full rate codes that perform very well in a variety of settings. The threaded algebraic space-time (TAST) codes \cite{GD:03}, codes constructed from cyclic division algebras (CDAs) \cite{SRS:03,BR:03}, and specifically modified CDA codes \cite{KR:05,EKP:06,ORB:06,ESK:07} that were shown (cf.\ \cite{EKP:06}) to achieve the entire DMT, are prime examples. The CDA based threaded designs are also the only currently known explicit constructions capable of achieving the DMT for all values of $\nt$ and simultaneously over all $r \in [0,\nt]$. All these codes have a common threaded structure.  Specifically an $n \times n$ threaded code is built from $n$ component codes mapped cyclically in threads (or layers) to the codewords $\bfX$. For example, in the special case of $n = \nt = T = 4$, the thread structure is given by
$$
\begin{bmatrix}
1 & 4 & 3 & 2 \\
2 & 1 & 4 & 3 \\
3 & 2 & 1 & 4 \\
4 & 3 & 2 & 1
\end{bmatrix}
$$
where the numbers $1,2,3,4$ indicate the thread to which a particular entry of $\bfX$ belongs.
In general, symbol $j$ in thread $l$ is mapped to $[\bfX]_{j,k}$ where
$
k = \mod\!( j-l \, , \, n) + 1
$
and where $\mod\!( \cdot \, , \, n)$ denotes the modulo $n$ operation. For example, in the case of perfect codes \cite{ORB:06,ESK:07} which also employ a threaded structure, the code follows from
\begin{equation} \label{eq:layers}
\lay(\bfX) = \theta \underbrace{\begin{bmatrix}
\bfB_0 \bfC & & \\
& \ddots & \\
& & \bfB_{n-1} \bfC
\end{bmatrix}}_{\bfUpsilon} \underbrace{\begin{bmatrix} \bfs^{(1)} \\ \vdots \\ \bfs^{(n)} \end{bmatrix}}_{\bfs}
\end{equation}
where
$$
\bfB_i \defeq \Diag(\underbrace{1,1,\ldots,1}_{\text{$n-i$ entries}},\underbrace{\gamma,\gamma,\ldots,\gamma}_{\text{$i$ entries}}),\ \ \ i=0,\cdots,n-1
$$
are full rank diagonal matrices incorporating a properly chosen thread-separating scalar $\gamma \in \complex$, where $\bfC \in \complex^{n \times n}$ is a (unitary) full rank generator matrix for the component code of each thread, $\bfs^{(l)} \in \const^n$ are the constellation symbols of thread $l$, and where $\lay(\bfX)$ denotes the matrix to vector operation obtained by stacking the elements of $\bfX$ according to their thread (cf.\ the column based stacking of the $\vec(\cdot)$ operation).

Regarding the cost of decoding by such codes, we note that the corresponding generator matrix $\bfG \in \complex^{n^2 \times n^2}$ is related to $\bfUpsilon$ through a permutation of the rows (cf.\ \eqref{eq:layers}) in such a way that the $(i,j)$th block $\bfG_{ij} \in \complex^{n \times n}$ of $\bfG$ contains exactly one non zero row which it self is one of the rows of $\bfB_{j-1}\bfC$. Consequently, $\bfG_{|ip} \defeq \begin{bmatrix} \bfG_{i1} & \cdots & \bfG_{ip} \end{bmatrix} \in \complex^{n \times np}$ has rank $p$ and contains exactly $p$ non-zero rows. This holds for any $n$ and $p\leq n$.  Now, let $\bfU_p \in \complex^{n \times p}$ be a unitary matrix with the property that any $p$ rows of $\bfU_p$ are linearly independent. Such matrices can clearly be constructed, and an example is the matrix that contains the first $p$ discrete Fourier transform (DFT) vectors of length $n$. Let $\bfGt_{|ip} \in \complex^{p \times np}$ be the matrix containing only the non-zero rows of $\bfG_{|ip} \in \complex^{p \times np}$, and let $\bfUt_{ip} \in \complex^{p \times p}$ be the full rank matrix consisting of the rows of $\bfU_p$ matching the non-zero rows of $\bfG_{|ip}$. It follows that
$$
(\bfI_n \kron \bfU_p\hr) \bfG_{|p} = \underbrace{\begin{bmatrix}
\bfUt_{1p}\hr & & \\
& \ddots & \\
& & \bfUt_{np}\hr
\end{bmatrix}}_{\bfUt_p} \underbrace{\begin{bmatrix}
\bfGt_{|1p} \\
\vdots \\
\bfGt_{|np}
\end{bmatrix}}_{\bfGt_{|p}} \in \complex^{np \times np}
$$
is full rank as both $\bfUt_p \in \complex^{np \times np}$ and $\bfGt_{|p} \in \complex^{np \times np}$ are full rank and square matrices. Note also that the same argument can be made regardless of the ordering of the threads, and for any other code with a threaded structure, provided the symbols in $\bfs$ are grouped into layers as in \eqref{eq:layers}. This is stated in the following.
\vspace{3pt}
\begin{theorem} \label{thrm:layered}
The SD complexity exponent, given any threaded code with $n=\nt=T$ that is decoded with the natural column ordering or under any other threat-wise grouping, is $c(r) = \bar{c}(r)$ where $\bar{c}(r)$ is given in Theorem \ref{thrm:upper-bound}.
\end{theorem}
\vspace{3pt}
Consequently directly from Theorems~\ref{thrm:upper-boundDMTopt} and \ref{thrm:layered}, we have the following result for DMT optimal threaded codes.

\vspace{3pt}
\begin{theorem} \label{thrm:layeredDMT}
Sphere decoding with thread-wise grouping of any DMT optimal threaded code with $n=\nt=T$, achieves DMT optimality with a SD complexity exponent of
\begin{equation} \label{eq:layeredDMT}
c(r) = r(n-\lfloor r \rfloor-1) + \big( n \lfloor r \rfloor - r(n-1) \big)^+ \,
\end{equation}
which, for integer values of $r=k$, simplifies to
\begin{equation} \label{eq:layeredDMTInteger}
c(k) = k(n-k).
\end{equation}
\end{theorem}
\vspace{3pt}

We briefly note that as expected the complexity increases with increasing $n = \nt = T$ for any fixed $r$ which is quite natural as the size of the codebook $\Xset$ and the signal space dimension increase. One can however also note that $c(r)$ is independent of the number of receive antennas $\nr$ (provided $\nr \geq \nt$). This is specific to the DMT optimal behavior and threaded structure of the codes, and may be explained by the fact that even though the channel quality generally improves by adding receive antennas - thus generally reducing complexity - the same improvement also occurs in the error probability performance of the code, and these two effects cancel each other in the SD complexity exponent.

\subsubsection{$2 \times 2$ approximately universal codes}

We here go one step further and identify a class of codes for which we can state, without limitations on the actual code structure, that $c(r) = \bar{c}(r)$ for any column ordering.  In particular, we establish this for the class of all $2\times 2$ approximately universal codes, i.e., all minimum delay codes that can achieve DMT optimality over the $2\times \nr$ channel irrespective of fading statistics~\cite{TV:06}. This is accomplished, albeit only for the specific case of $\nt \times T = 2 \times 2$, by proving that \eqref{eq:rank-condition} follows from the so called non-vanishing determinant (NVD) condition \cite{BR:03} which is well known to be a necessary and sufficient condition for approximate universality.  We consequently have the following.

\vspace{3pt}
\begin{theorem} \label{thrm:nvd}
Any $2 \times 2$ full rate approximately universal linear dispersive code, irrespective of its structure, introduces a SD complexity exponent of
$$
c(r) = \min(r,2-r).
$$
\end{theorem}
\vspace{3pt}

As the NVD property does not depend on the ordering of the columns of $\bfG$, it also follows that the conclusion of Theorem~\ref{thrm:nvd} holds irrespective of the column ordering.

\section{Implications and Discussions} \label{sec:discussions}

\subsection{Decoding complexity and information theoretic outages} \label{sec:outages}

\begin{figure*}
\psfrag{a}[cc]{\small 0}
\psfrag{b}[cc]{\small 1/2}
\psfrag{c}[cc]{\small 1}
\psfrag{d}[cc]{\small 3/2}
\psfrag{0}[cc]{\small 0}
\psfrag{1}[cc]{\small 1}
\psfrag{2}[cc]{\small 2}
\psfrag{3}[cc]{\small 3}
\psfrag{4}[cc]{\small 4}
\psfrag{r0.5}[lb]{$r = 1/2$}
\psfrag{r1.0}[lb]{$r = 1$}
\psfrag{r1.5}[lb]{$r = 3/2$}
\psfrag{ca}[bc]{$\bar{c}(r:\bfalpha)$}
\psfrag{a1}[cc]{$\alpha_1$}
\psfrag{a2}[cc]{$\alpha_2$}
\includegraphics[width=\linewidth]{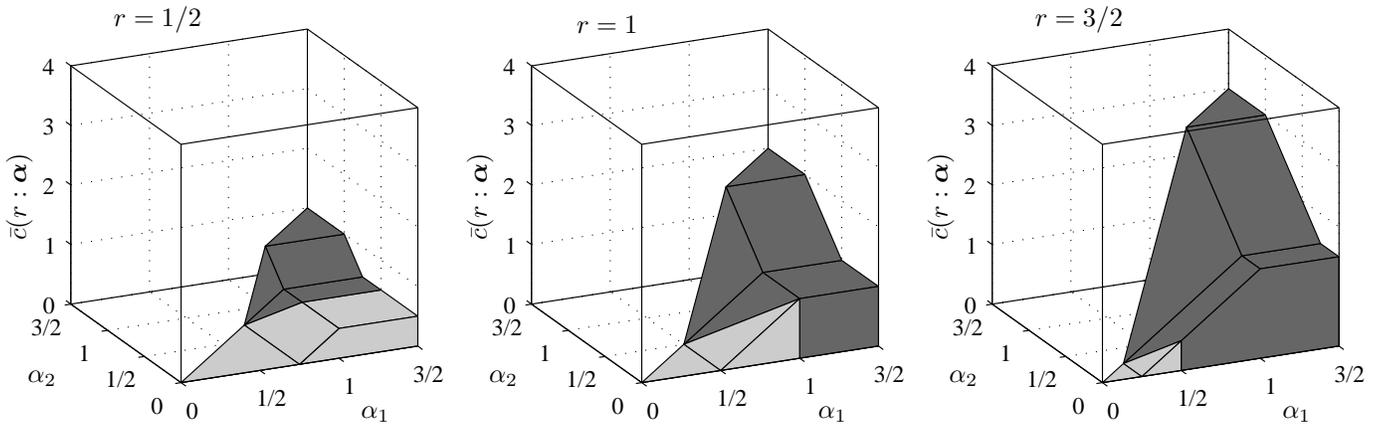}
\caption{Conditional SD complexity exponent bound $\bar{c}(\bfalpha)$ as a function of singularity levels $\bfalpha = (\alpha_1,\alpha_2)$. Singularity levels that corresponds to outage events at the target multiplexing gain $r$ are shown in dark grey, while singularity levels capable of supporting the target multiplexing gain are shown in light grey.}
\label{fig:outages}
\end{figure*}

We recall that the claim of Theorem \ref{thrm:upper-bound} may be expressed in terms of the function (cf.\ \eqref{eq:optimization-objective})
\begin{equation}
\label{eq:complexityForAlpha}\bar{c}(r:\bfalpha) \defeq \sum_{i=1}^{\nt} T \min\!
 \left( \frac{r}{\nt} - 1 + \alpha_{i}\, , \, \frac{r}{\nt} \right)^{\!+} \, ,
\end{equation}
which provides a conditional, asymptotic, upper bound on the sphere decoding complexity according to $N \dotleq \rho^{\bar{c}(r:\bfalpha)}$ in terms of the multiplexing gain $r$ and the singularity level $\bfalpha$. The final upper bound $\bar{c}(r)$ in \eqref{eq:optimization} is then given as the worst-case $\bar{c}(r:\bfalpha)$ over all singularity levels that occur with a probability that is larger than or equal to the probability of error, given asymptotically by the diversity gain $d(r)$ of the code.

The characterization of the DMT in \cite{ZT:03} relies on the asymptotic probability of outages at high SNR, i.e., the probability that the i.i.d.\ Rayleigh fading AWGN channel given by
$$
\bfy_t = \bfH \bfx_t + \bfw_t
$$
with a power constraint $\expt{\|\bfx_t\|^2} \leq \rho$ cannot support an asymptotic data-rate of $R = r \log \rho + o(\log \rho)$. As was shown in \cite{ZT:03}, this occurs when the singularity levels belong to the outage set $\Aset(r) = \{ \bfalpha \, | \, \sum_i (1-\alpha_i)^+ < r \}$, and the diversity of the outage event is given by the most likely set of singularity levels that satisfy this condition, i.e., $d(r) = \inf_{\bfalpha \in \Aset(r)} I(\bfalpha)$  where $I(\bfalpha)$ is given in \eqref{eq:rate-function}.

If we restrict attention to the set of singularity levels whose probability of occurring does not vanish exponentially fast, i.e., for which $I(\bfalpha) < \infty$ or equivalently $\alpha_\nt \geq \ldots \geq \alpha_1 \geq 0$, we can for DMT optimal codes\footnote{To be precise, we are assuming here that we are working with approximately universal codes for which it is known that errors are only likely when the channel is in outage~\cite{TV:06}.  All explicitly constructed full rate DMT-optimal codes known to date, are also approximately universal.} make an interesting connection between the decoding complexity and information theoretic outages. In particular, as $d(r) = \inf_{\bfalpha \in \Aset(r)} I(\bfalpha)$ it follows that $\sum_i (\nr - \nt + 2i-1)\alpha_i \leq d(r)$ if and only if $\sum_i (1-\alpha_i)^+ \geq r$. We can thus, for DMT optimal codes,
equivalently express \eqref{eq:optimization} according to
\begin{subequations} \label{eq:outage-interpretation}
\begin{align}
\bar{c}(r) \, = \, \max_{\bfalpha} \; & \bar{c}(r:\bfalpha) \\
\text{s.t.} \; &  \sum_{i=1}^{\nt} (1-\alpha_i)^+ \geq r \label{eq:outage-interpretation-c1} \\
 & \alpha_{1} \geq \ldots \geq \alpha_{\nt} \geq 0 \label{eq:outage-interpretation-c2} \, ,
\end{align}
\end{subequations}
which may be interpreted as the worst-case complexity (bound) over all channels that are not in outage. This significantly strengthens the connection between channel and decoding outages touched upon in Section \ref{sec:decoding-complexity}.

The concept is illustrated for $\nt=T=2$ in Fig.~\ref{fig:outages} where $\bar{c}(r:\bfalpha)$ is plotted as a function of $\bfalpha = (\alpha_1,\alpha_2)$ over $\alpha_1 \geq \alpha_2 \geq 0$. In this case, $\bar{c}(r) = \min(r,2-r)$. Note also here that we know by Theorem \ref{thrm:nvd} that $c(r) = \bar{c}(r)$. Singularity levels that are in the outage region $\sum_i (1-\alpha_i)^+ < r$ are shown in a darker shade. It can be seen that increasing the multiplexing gain $r$ increases the codeword density and codebook size and consequently broadens the set of singularity levels that can potentially lead to higher complexity. However, increasing the multiplexing gain also reduces the set of channels that support the data-rate, thus limiting the set of singularity levels for which the decoder needs to be applied, leading to an overall reduction in the SD complexity exponent as $r$ approaches its maximum value.

Further, the connection to information theoretic outages allows for an intuitive explanation of the result of Theorem~\ref{thrm:layeredDMT}, and in particular \eqref{eq:layeredDMTInteger}. To this end, it is illustrative to consider a heuristic argument involving low rank channel matrices $\bfH$. As noted in \cite{ZT:03}, the typical outages at integer multiplexing gains $r=k$ are caused by channels that are close to the set of rank $k$ matrices, i.e., that have $n-k$ small singular values. If we for the purpose of illustration assume that $\bfH$ has rank $k$, it follows that $\bfI_T \kron \bfH$ for $T=n$, and $\bfM$, has rank $nk$, and consequently a null-space of dimension $n(n-k)$. This implies that the $n(n-k) \times n(n-k)$ lower right block of $\bfR$ is\footnote{This also requires that the first $nk$ columns of $\bfR$ are linearly independent.
In fact, the rigorous treatment of this technical detail is largely responsible for much of the difficulty in establishing the lower bounds on $c(r)$. In particular, condition \eqref{eq:rank-condition} in Lemma \ref{lm:lower-bound} guarantees that this happens with sufficiently high probability, while Lemma \ref{lm:perturbation} provides a perturbation analysis that allows us to extent this intuitive reasoning to channels that are close to the set of rank deficient matrices.} identically equal to zero, and the sphere decoder pruning criteria become totally ineffective up to and including layer $n(n-k)$. As the size of $\const$ is $|\const| \doteq \rho^{\frac{k}{n}}$ for $r=k$, we see that the number of nodes searched at layer $n(n-k)$ of the SD search tree is $|\const^{n(n-k)}| \doteq \rho^{k(n-k)}$ (cf.\ \eqref{eq:layeredDMTInteger}). In order to ensure close to optimal performance, the sphere decoder must be able to decode for channels where $n-k$ singular values are close to zero. However, channels with even more singular values close to zero occur with a probability that is small in relation to the outage probability or the probability of ML decoder error, and can thus be safely ignored by the decoder.

\subsection{A complexity bound that holds for all fading statistics} \label{sec:other-fading}

It is perfectly conceivable that the sphere decoding complexity may rise under specific codes and under specific fading statistics that tend to regularly introduce channel instances that are difficult to decode for. A natural question is then whether one can bound the complexity, irrespective of the code and of the statistical characterization of the channel. Considering Theorem~\ref{thrm:upper-bound}, and the proof of this theorem, we can see that the i.i.d.~Rayleigh fading assumption only enters through the rate function $I(\bfalpha)$. Consequently, we may directly restate Theorem~\ref{thrm:upper-bound} for other fading distributions after updating \eqref{eq:optimization-div} and \eqref{eq:optimization-pos} with the appropriate rate-function $I(\bfalpha)$. Some relevant examples of rate-functions for other fading distributions are given in \cite{ZMM:07}.

Further, regardless of which $I(\bfalpha)$ applies, the upper bound $\bar{c}(r)$ in Theorem~\ref{thrm:upper-bound} is non-decreasing in $d(r)$ and maximized when $d(r)$ corresponds to the outage exponent. In this case $\bar{c}(r)$ is, again, given by \eqref{eq:outage-interpretation}, which does not explicitly depend the fading distribution other than through the assumption that $P(\alpha_\nt<0)$ vanishes exponentially fast; an assumption that holds for all reasonable distributions. This implies that SD complexity exponent is universally upper bounded as (cf.\ Theorem~\ref{thrm:upper-boundDMTopt})
\[c(r)\leq \frac{T}{\nt} \Big( r(\nt-\lfloor r \rfloor-1) + \big( \nt \lfloor r \rfloor - r(\nt-1) \big)^+ \Big) \, \]
for any full-rate code and statistical characterization of the channel.
This is also clearly the tightest upper bound that can hold for all (full-rate) codes and fading statistics.

\subsection{Fast decodable codes}

In \cite{TK:02,PGG:07,SF:07} a family of DMT optimal $\nt \times T = 2 \times 2$ space-time codes called fast decodable codes \cite{BHV:09} were constructed. The SD complexity exponent (and also its upper bound) provides an interesting approach for comparing the complexity of decoding regular codes and fast decodable codes. Before doing so it should be noted that these fast decodable codes are not, strictly speaking, of the form in \eqref{eq:ld-form} as the real and imaginary part of each constituent symbol is dispersed separately. Nevertheless, the fast decodable codes may be decoded by an equivalent \emph{real valued} sphere decoder that performs a search over a tree with $2\kappa$ layers and $|\const|^{\frac{1}{2}}$ branches per node, and we can compare the reported worst-case complexity of this real valued sphere decoder to the complexity of the complex valued sphere decoder considered herein.

The fast decodable codes have the appealing property that the upper right $4 \times 4$ block of the real valued $\bfR \in \reals^{8 \times 8}$ (cf.\ \eqref{eq:sd-metric}) is always a diagonal matrix, regardless of the particular realization of $\bfH$. While the regular real valued sphere decoder for a $2 \times 2$ full rate code would perform a (bounded) search over the entire tree, it is sufficient for the fast decodable codes to (without loss of optimality) perform a search over only the $4$ first layers, and extend each node at layer $4$ to a valid codeword through a faster, linear, ML decoding. This simplified version of the real valued sphere decoder can be viewed as a search over a regular tree where each node has $|\const|^{\frac{1}{2}}$ children up to layer $4$, but only one child per node for the $4$ remaining layers. Consequently, the worst-case number of nodes visited by the simplified sphere decoder is $5|\const|^{\frac{4}{2}} + \sum_{k=1}^{3} |\const|^{\frac{k}{2}} \doteq |\const|^{2} \doteq \rho^{r}$ as opposed to $\sum_{k=1}^{8} |\const|^{\frac{k}{2}} \doteq \rho^{2r}$ for the regular real valued sphere decoder, cf.\ \cite{BHV:09}. Thus, fast decodability implies a reduction by a factor of 2 in the worst-case SNR exponent, which is significant at high SNR.

\begin{figure}
\begin{center}
\psfrag{r1}[bc]{$r$}
\psfrag{r2}[bc]{$2r$}
\psfrag{ce}[bc]{$c(r)$}
\psfrag{cey}[bc]{SNR exponent}
\psfrag{r}[c]{Multiplexing gain $r$}
\psfrag{0}{\small 0}
\psfrag{1}{\small 1}
\psfrag{2}{\small 2}
\psfrag{3}{\small 3}
\psfrag{4}{\small 4}
\psfrag{0.5}{\small $1/2$}
\psfrag{1.5}{\small $3/2$}
\includegraphics[width=8.3cm]{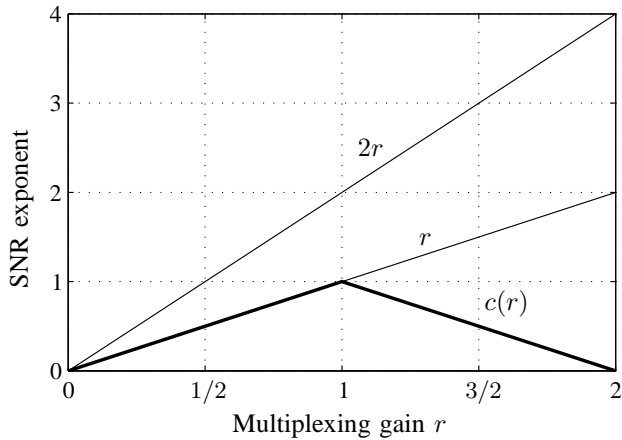}
\end{center}
\caption{Comparing the SD complexity exponent $c(r) = \min(r,2-r)$, with the (worst-case) SNR exponent $(2r)$ of the regular sphere decoder, and with the maximal SNR exponent $(r)$ of the simplified sphere decoder.} \label{fig:fastdec}
\end{figure}

However, this worst-case SNR exponent of the simplified SD algorithm should be viewed in light of the SD complexity exponent induced by any $2 \times 2$ approximately universal code as given by Theorem \ref{thrm:nvd}, i.e., $c(r) = \min(r,2-r)$. The worst-case SNR exponent of the regular sphere decoder and of the simplified sphere decoder, and the SD complexity exponent $c(r)$ are shown in Fig.~\ref{fig:fastdec}. For multiplexing gains lower than or equal to $1$, the SD complexity exponent and the worst-case SNR exponent of the simplified sphere decoder actually coincide, while the SD complexity exponent is strictly lower for multiplexing gains higher than one. The interpretation is that a run-time constrained sphere decoder will yield asymptotic ML performance for \emph{any} $2 \times 2$ approximately universal code at a complexity that is comparable to the reported worst-case complexity of the fast decodable codes at low multiplexing gains, and significantly better at high multiplexing gains. Thus, in a sense, all approximately universal codes are fast decodable at high SNR. However, we hasten to add that the fast decodable structure can naturally still be desirable in many cases of practical interest.

\subsection{The applicability of Lemma~\ref{lm:lower-bound}}

Finally, we discuss the application of Lemma~\ref{lm:lower-bound} to codes not considered herein. To this end, note that for any given generator matrix $\bfG$ of some code not covered by Section \ref{sec:tightness}, it should be clear that if \eqref{eq:rank-condition} holds then Lemma~\ref{lm:lower-bound} could be used to establish a tight lower bound on $c(r)$. This said, we also wish to caution the reader that \eqref{eq:rank-condition} only represents a sufficient condition for $c(r) = \bar{c}(r)$. It does not necessarily follow that $c(r) < \bar{c}(r)$ if \eqref{eq:rank-condition} is not true. In other words, the question of if there are code designs that improve upon the bound $\bar{c}(r)$ is not answered in the positive by finding code designs for which \eqref{eq:rank-condition} does not hold.

As for testing \eqref{eq:rank-condition} it should also be noted that one does not have to restrict the search for $\bfU_p$ to the set of unitary matrices. Any full rank matrix $\bfA \in \complex^{\nt \times p}$ can be factored, e.g., by the QR decomposition, as $\bfU_p \bfT = \bfA$ where $\bfT \in \complex^{p \times p}$ has rank $p$ and where $\bfU_p$ is unitary. Hence, as $(\bfI_T \kron \bfT\hr)$ is full rank, it follows that
$$
(\bfI_T \kron \bfA\hr) \bfG_{|p} = (\bfI_T \kron \bfT\hr) (\bfI_T \kron \bfU_p\hr) \bfG_{|p}
$$
is rank deficient if and only if \eqref{eq:rank-condition} fails to hold. Hence, the statement of Lemma~\ref{lm:lower-bound} could be phrased in terms of the existence of any $\bfU_p$, not necessarily unitary.

Further, let
$$
p(\bfA) \defeq |(\bfI_T \kron \bfA\hr) \bfG_{|p}| \, ,
$$
where $|\cdot|$ denotes the determinant, and note that $p(\bfA)$ is a polynomial in the elements of $\bfA$. It can thus be seen that if $p(\bfA) \neq 0$ for some $\bfA \in \complex^{\nt \times p}$, i.e., $p(\bfA)$ is not the zero polynomial, it follows that the set of $\bfA$ for which $p(\bfA) = 0$ has zero Lebesgue measure. It is then a straightforward extension to show that \eqref{eq:rank-condition} holds either never or almost always with respect to the set of unitary matrices $\bfU_p$ over the Stiefel manifold (i.e., the set of all unitary $n \times p$ matrices) endowed with the Haar (uniform) measure. This suggests a rather interesting conceptual method for verifying \eqref{eq:rank-condition}. Given a specific generator matrix one could at least in theory test the condition of Lemma \ref{lm:lower-bound} by selecting $\bfU_p$ (or $\bfA$) uniformly at random, and the condition of Lemma \ref{lm:lower-bound} would be proven with probability one if true. However, finite precision computations will limit the practical applicability of such an approach, although symbolic computations could potentially be a way to test a specific code design.

\section{Conclusion}
\label{sec:conclusion}

The work addressed the open question of identifying the computational cost of near-ML sphere decoding.  In the high-SNR high-rate regime, the introduced SD complexity exponent asymptotically described this cost, concisely revealing the cost's natural dependencies to the codeword density, the codebook size, as well as to the SNR, dimensions and fading characteristics of the wireless channel. This exponent currently sets the bar with respect to the computational reserves required for decoding with arbitrarily close to ML performance, and the clear challenge is now to identify transceivers with a lesser complexity exponent that can still guarantee a vanishing ML gap.

The simplicity of the provided guarantees can offer insight into designing robust encoders, decoders, and time-out policies, as well as guidelines for network planning in settings where rate, reliability, and computational complexity are principal intertwined concerns.  Such guarantees can apply towards substantial savings in energy, processing power, and hardware.

\begin{appendices}

\section{Proof of Lemma 1} \label{app:upper-bound}

\begin{figure}
\begin{center}
\psfrag{c1}[c]{$\Cset_{1}$}
\psfrag{c2}[c]{$\Cset_{2}$}
\psfrag{c3}[c]{$\Cset_{3}$}
\psfrag{c4}[c]{$\Cset_{4}$}
\psfrag{e}[c]{$\Eset$}
\psfrag{b}[c]{$\Bset$}
\includegraphics[width=8.3cm]{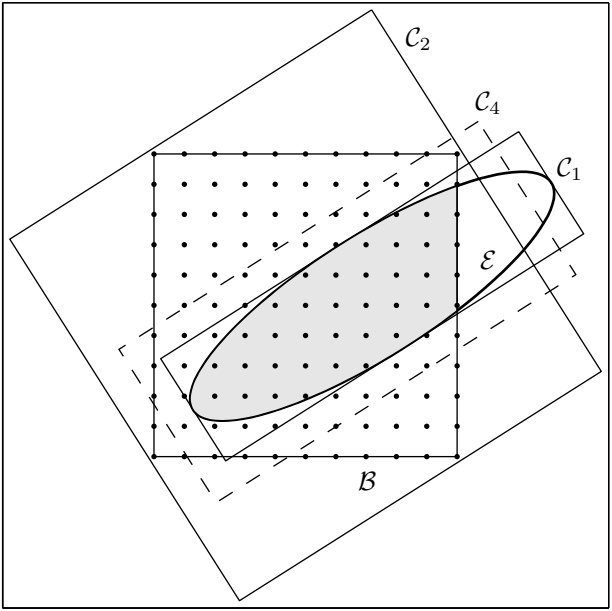}
\end{center}
\caption{Illustration of the proof of \eqref{eq:lm-upper} in Lemma \ref{lm:upper-bound} in the case of $n=2$. The lemma provides an upper bound on the number of integer points within the shaded area, corresponding to the intersection of the ellipsoid and the constellation boundary.} \label{fig:lattice}
\end{figure}
In the following, we provide a proof of Lemma \ref{lm:upper-bound}, starting with the upper bound in \eqref{eq:lm-upper} and then establishing the lower bound in \eqref{eq:lm-lower}. To this end, note that the length of the $i$th semi-axis of $\Eset$, denoted $e_{i}$, is given by
$$
e_{i} \defeq  \frac{2\xi}{\sigma_{i}(\bfD)} \, .
$$
Let $\Cset_{1}$ be the smallest orthotope (box), aligned with and containing $\Eset$, i.e., $\Cset_{1}$ is an orthotope with side lengths $e_i$ (see Fig.\ \ref{fig:lattice}). Let $\Cset_{2}$ be a hypercube with side-length $2\sqrt{n}\eta$, centered at the origin and aligned with $\Cset_{1}$ (see Fig.\ \ref{fig:lattice}). As the diagonal of $\Bset$ is $2\sqrt{n}\eta$ it follows that $\Bset \subset \Cset_{2}$, regardless of the orientation of $\Cset_{2}$. Let $\Cset_{3}$ be given by $\Cset_{3} = \Cset_{1} \cap \Cset_{2}$ and note that $\Eset \cap \Bset \subset \Cset_{3}$ as $\Eset \subset \Cset_{1}$ and $\Bset \subset \Cset_{2}$. As $\Cset_{1}$ and $\Cset_{2}$ are aligned, it follows that $\Cset_{3}$ is also an orthotope. Let $l_{1},\ldots,l_{n}$ denote the side-lengths of $\Cset_{3}$ and note that $l_{i} \leq \min( e_{i} \, , \, 2\sqrt{n}\eta)$.

The mean value theorem \cite{GW:93} states that for any convex body (or set) $\Cset \subset \reals^n$ it holds that
\begin{equation} \label{eq:pointvolume}
\text{Vol}(\Cset) = \int_{\Uset}  |\ints^n \cap \Cset + \bfu | d\bfu
\end{equation}
where
$$
\Uset \defeq \big[-\tfrac{1}{2},\tfrac{1}{2} \big]^n
$$
denotes the unit cube in $\reals^n$, i.e., the volume of the set equals
the average number of integer points in the set when perturbed by a uniform random perturbation over the unit cube \cite{GW:93}. This statement is a rigorous version of the intuitive notion that the number of integer points in $\Cset$ may be approximated by its volume, and it can be used to obtain a non-random upper bound on the number of integer points in $\Cset_3$. To this end, let $\Cset_{4}$ be the orthotope, aligned with and centered around $\Cset_{3}$, with side lengths $l_{i}+\sqrt{n}$ (see Fig.\ \ref{fig:lattice}). By construction, it follows that
$$
\Cset_{3} \subset \Cset_{4} + \bfu
$$
for \emph{any} $\bfu \in \Uset$. It therefore follows by \eqref{eq:pointvolume} that
$$
|\Cset_{3} \cap \ints^n| \leq \text{Vol}(\Cset_{4}) = \prod_{i=1}^n \left[ \sqrt{n}+ l_{i} \right] \, ,
$$
where
$$
l_{i} \leq \min\left( \frac{2 \xi}{\sigma_{i}(\bfD)} \, , \, 2 \sqrt{n} \eta \right ) \, .
$$
As $\Eset \cap \Bset \subset \Cset_{3}$ the upper bound in \eqref{eq:lm-upper} follows.

In order to establish the lower bound in \eqref{eq:lm-lower} we may redefine $\Cset_{1}$ to be the orthotope, aligned with the semiaxes of $\Eset$ and with the same center, having side-lengths
$$
b_{i} = \frac{2\xi}{\sqrt{n}\sigma_{i}(\bfD)} \, ,
$$
for $i=1,\ldots,n$. Now, by construction $\Cset_{1} \subset \Eset$ which implies that $|\Cset_{1} \cap \ints^n| \leq |\Eset \cap \ints^n|$. Let $\Cset_{2}$ be another orthotope, aligned with $\Cset_{1}$ and with side-lengths $\max(b_{i}-\sqrt{n} \, , \, 0)$. It follows that $\Cset_{2} + \bfu \subset \Cset_{1}$ for any $\bfu \in \Uset$. By reasoning similar to what is used in the proof of the upper bound (cf.\ \eqref{eq:pointvolume}) it follows that
$$
| \Eset \cap \ints^n | \geq \mathrm{vol}(\Cset_{2}) = \prod_{i=1}^n \max(b_{i}-\sqrt{n} \, , \, 0) \, ,
$$
which establish the lower bound in \eqref{eq:lm-lower}.

\section{Proof of Lemma 2} \label{app:lower-bound}

Let $\bfalpha^\star = (\alpha_{1}^\star, \ldots, \alpha_{\nt}^\star)$ be an optimal point of \eqref{eq:optimization} and let $q$ be the largest integer for which (cf.\ \eqref{eq:optimization-objective})
\begin{equation} \label{eq:pos-constraint}
\frac{r}{\nt} - 1 + \alpha_{q}^\star > 0 \, .
\end{equation}
Note that we can without loss of generality assume that $q \geq 1$ as otherwise $\bar{c}(r) = 0$ (cf.\ \eqref{eq:optimization-objective}) and $c(r) = \bar{c}(r)$ would be a trivial statement.
It follows that $\alpha^\star_i > 0$ for $i=1,\ldots,q$ and we may also without loss of generality assume that $\alpha^\star_{i} \leq 1$ for $i=1,\ldots,\nt$ as the objective in \eqref{eq:optimization} does not increase in $\alpha_i$ beyond $\alpha_i = 1$. The goal will be to show that layer $k = qT$ of the sphere decoder contains close to $\rho^{\bar{c}(r)}$ nodes with a probability that is large with respect to the probability of decoding error $\prob{\bfX_\ml \neq \bfX} \doteq \rho^{-d(r)}$.

To this end, let $\bfH = \bfU \bfSigma \bfV\hr$ be the singular value decomposition of $\bfH$, where
$$
\bfSigma \defeq \Diag(\sigma_{1}(\bfH), \ldots, \sigma_{\nt}(\bfH))
$$
and where $\bfU\hr\bfU = \bfI$. Let $\bfU_{p}$ denote the last $p \defeq \nt-q$ columns of $\bfU$ (corresponding to the $p$ largest singular values) and let $\bfalpha = (\alpha_{1},\ldots,\alpha_{\nt})$ be the random vector of singularity levels given by \eqref{eq:eigenvalue-asymptot}. Now, consider the set of conditions (or events) given by
\begin{subequations} \label{eq:conditions}
\begin{align} \label{eq:condition-alpha}
\Omega_{1} \defeq \, \{ \, & \alpha_{i}^\star - 2\delta < \alpha_{i} < \alpha_{i}^{\star} - \delta \, , \; i=1,\ldots,q \, , \nonumber \\
& 0 < \alpha_{i} < \delta \, , \; i=q+1,\ldots,\nt \} \, ,
\end{align}
for some given (small) $\delta > 0$,
\begin{equation} \label{eq:condition-u}
\Omega_{2} \defeq \{ \sigma_{1}((\bfI_{T} \kron \bfU_p\hr) \bfG_{|p}) \geq u \} \, ,
\end{equation}
for some given $u > 0$,
\begin{equation} \label{eq:condition-w}
\Omega_{3} \defeq \{ \|\bfQ\hr\bfw\| \leq 1 \} \, ,
\end{equation}
and
\begin{equation} \label{eq:condition-s}
\Omega_{4} \defeq \{ \|\bfs\| \leq \tfrac{1}{2} \eta \} \, .
\end{equation}
\end{subequations}
Note also that by choosing $\delta$ sufficiently small, we may without loss of generality assume that $\Omega_1$ implies that $\alpha_i > 0$ for all $i=1,\ldots,\nt$.

The following proof is structured as follows: First, in Sections \ref{sec:boundary-condition} and \ref{sec:eigenvalue-bound}, it is established that \eqref{eq:conditions} represent sufficient conditions for the number of nodes $N_{k}$ visited in layer $k=qT$ to be close to $\rho^{\bar{c}(r)}$. Then, in Section \ref{sec:probability}, it is established that the set of conditions in \eqref{eq:conditions} are simultaneously satisfied with a probability that is large with respect to the probability of error.

\subsection{The Constellation boundary} \label{sec:boundary-condition}

We begin by proving that, given \eqref{eq:conditions}, the constellation boundary may be ignored, i.e., that $\const$ may be replaced by $\constme_\infty$ in \eqref{eq:nodes-per-layer} without changing the result, thus making the lower bound \eqref{eq:lm-lower} in Lemma \ref{lm:upper-bound} applicable. To this end, let $\bfsh_{k} \in \constme_\infty^k$ be an arbitrary point in the $k$-dimensional infinite constellation (i.e., the Gaussian integer lattice) and assume that $\bfsh_{k}$ satisfies the sphere constraint at layer $k$, i.e.\
$$
\| \bfr_{k} - \bfR_{k} \bfsh_{k} \| \leq \xi \, .
$$
Note that $\bfr_{k} = \bfR_{k} \bfs_{k} + \bfv_{k}$, where $\bfs_{k}$ denotes the last $k$ components of the transmitted symbol vector $\bfs \in \const^\kappa$ and where $\bfv_{k}$ denotes the last $k$ components of $\bfv \defeq \bfQ\hr\bfw$. It follows that
\begin{align*}
\| \bfr_{k} - \bfR_{k} \bfsh_{k} \| = & \, \| \bfR_{k} (\bfs_{k} - \bfsh_{k} ) + \bfv_{k} \bf \| \\
\geq & \, \sigma_{1}(\bfR_{k}) \| \bfsh_{k} - \bfs_{k}\| - \| \bfv_{k} \|
\end{align*}
which implies that
$$
\|\bfsh_{k} - \bfs_{k} \| \leq \frac{1}{\sigma_{1}(\bfR_{k})} (\xi + \|\bfv_{k}\|)
$$
and
\begin{equation} \label{eq:symbols-bound}
\|\bfsh_{k} \| \leq \frac{1}{\sigma_{1}(\bfR_{k})} (\xi + \|\bfv_{k}\|) + \|\bfs_{k}\| \, .
\end{equation}
By the interlacing property of singular values (cf.\ \eqref{eq:interlace}) it further follows that
$$
\sigma_{1}(\bfR_{k}) \geq \theta \gamma \sigma_{1}(\bfH) \doteq \rho^{\frac{1}{2}-\frac{rT}{2\kappa}-\frac{1}{2}\alpha_{1}} \geq \rho^{\frac{1}{2}\delta -\frac{rT}{2\kappa} }
$$
where we recall that $\theta \doteq \rho^{\frac{1}{2}-\frac{rT}{2\kappa} }$ is the power scaling and $\gamma = \sigma_1(\bfG) > 0$, and where the last inequality is implied by \eqref{eq:condition-alpha} and $\alpha_{1}^\star \leq 1$. As $\xi \doteq \rho^0$ and $\| \bfv_{k} \| \leq \| \bfQ\hr\bfw \| \leq 1$ by \eqref{eq:condition-w} it follows that
$$
\frac{1}{\sigma_{1}(\bfR_{k})} (\xi + \|\bfv_{k}\|) \dotleq \rho^{\frac{rT}{2\kappa} - \frac{1}{2}\delta} \, .
$$
By \eqref{eq:symbols-bound}, \eqref{eq:condition-s}, $\|\bfs_{k}\| \leq \|\bfs\| \leq \eta$ and since $\rho^{\frac{rT}{2\kappa} - \frac{1}{2}\delta} \dotle \tfrac{1}{2}\eta$, it follows that
$$
\| \bfsh_{k} \| \leq \eta
$$
given that $\rho$ is sufficiently large. This implies that $\bfsh_{k} \in \const^k$. Thus, any integer point that satisfies the sphere constraint must also belong to the constellation, and we can proceed using \eqref{eq:lm-lower} to lower bound the complexity.

\subsection{Singular value bounds} \label{sec:eigenvalue-bound}

We proceed to provide bounds on the singular values of $\bfR_{k}$ in order to lower bound the number of nodes visited in layer $k=qT$. However, as stated previously the interlacing theorem is, unlike in the derivation of the upper bound, not sufficient for our purposes. Instead, we consider the following lemma, proven in Appendix \ref{app:perturbation}.

\vspace{5pt}
\begin{lemma} \label{lm:perturbation}
Let $\bfA \in \complex^{m \times n}$, $m \geq n$ be an arbitrary matrix and $\bfQ \bfR = \bfA$ be the QR decomposition of $\bfA$. Partition $\bfA$, $\bfQ$ and $\bfR$ according to
$$
\begin{bmatrix}\bfA_{1} & \bfA_{2} \end{bmatrix} =
\begin{bmatrix}\bfQ_{1} & \bfQ_{2} \end{bmatrix} \begin{bmatrix} \bfR_{11} & \bfR_{12} \\ \bfZero & \bfR_{22}
\end{bmatrix} \, .
$$
where $\bfA_{1} \in \complex^{m \times n-k}$ and $\bfR_{22} \in \complex^{k \times k}$. Then, assuming that $\sigma_{i}(\bfA) < \sigma_{1}(\bfA_{1})$ for $i=1,\ldots,k$, it holds that
\begin{equation} \label{eq:perturbationbound}
\sigma_{i}(\bfR_{22}) \leq \left[ \frac{\sigma_{n}(\bfA)}{\sigma_{1}(\bfA_{1})} + 1 \right] \sigma_{i}(\bfA) \, .
\end{equation}
\end{lemma}
\vspace{9pt}

Applied to the effective channel matrix $\bfM$ it follows that
\begin{equation} \label{eq:perturbationbound-2}
\sigma_{i}(\bfR_k) \leq \left[ \frac{\sigma_{\kappa}(\bfM)}{\sigma_{1}(\bfM_{1})} + 1 \right] \sigma_{i}(\bfM)
\end{equation}
where $\bfM_{1}$ contains the first $pT$ columns ($p=\nt-q$) of $\bfM$, assuming that $\sigma_{i}(\bfM) < \sigma_{1}(\bfM_{1})$ as will be shown for $i=1,\ldots,qT$ later. In order to lower bound $\sigma_{1}(\bfM_{1})$ note that
$$
\bfM_{1} = \theta (\bfI_{T} \kron \bfH) \bfG_{|p} \, ,
$$
where $\bfG_{|p}$ denotes the first $pT$ columns of $\bfG$ and
$$
\bfM_{1}\hr\bfM_{1} = \theta^2 \bfG_{|p}\hr(\bfI_{T} \kron \bfH\hr\bfH) \bfG_{|p} \, .
$$
As
$$
\bfH\hr\bfH \succeq \sigma_{q+1}(\bfH\hr\bfH) \bfU_{p}\bfU_{p}\hr
$$
where $\bfU_p$ denotes the matrix containing the $p$ singular vectors corresponding to the $p$ largest singular values, and where $\bfA \succeq \bfB$ denotes that $\bfA-\bfB$ is positive semi-definite, it follows that
$$
\bfM_{1}\hr\bfM_{1} \succeq \theta^2 \sigma_{q+1}(\bfH\hr\bfH) \bfG_{|p}\hr(\bfI_{T} \kron \bfU_{p}\bfU_{p}\hr) \bfG_{|p} \, .
$$
Considering the smallest singular value of $\bfM_{1}\hr\bfM_{1}$ yields
$$
\sigma_{1}(\bfM_{1}\hr\bfM_{1}) \geq \theta^2 \sigma_{q+1}(\bfH\hr\bfH) \sigma_{1}(\bfG_{|p}\hr(\bfI_{T} \kron \bfU_{p}\bfU_{p}\hr) \bfG_{|p})
$$
and concequently
\begin{equation} \label{eq:sing-bound-1}
\sigma_{1}(\bfM_{1}) \geq u \theta \sigma_{q+1}(\bfH) = u \theta \rho^{-\frac{1}{2}a_{q+1}} \dotgeq \rho^{\frac{1}{2}-\frac{rT}{2\kappa}-\frac{1}{2} \delta} \, ,
\end{equation}
where the first inequality follows by \eqref{eq:condition-u} and the last inequality follows by \eqref{eq:condition-alpha} together with $\theta \doteq \rho^{\frac{1}{2}-\frac{rT}{2\kappa}}$ and as $u > 0$ is fixed (independent of $\rho$). Further,
$$
\sigma_{i}(\bfM) = \theta \sigma_i((\bfI_T \kron \bfH)\bfG) \leq \theta \Gamma \sigma_{\iota_{T}(i)}(\bfH)
$$
where $\Gamma \defeq \sigma_{\max}(\bfG) = \sigma_{\kappa}(\bfG)$, and it follows from \eqref{eq:condition-alpha} that
\begin{equation} \label{eq:sing-bound-2}
\sigma_{i}(\bfM) \dotleq \rho^{\frac{1}{2}-\frac{rT}{2\kappa}-\frac{1}{2}\alpha^\star_{\iota_{T}(i)} + \delta}
\end{equation}
for $i=1,\ldots,qT$. As $\alpha^\star_{i} > 0$ for $i=1,\ldots,q$, it follows by comparing \eqref{eq:sing-bound-1} and \eqref{eq:sing-bound-2} that $\sigma_{i}(\bfM) \leq \sigma_{1}(\bfM_{1})$ for $i=1,\ldots,qT$, given that $\delta$ is sufficiently small and that $\rho$ is sufficiently large, making Lemma \ref{lm:perturbation} applicable for $k=qT$.

For the maximal singular value of $\bfM$ we have (cf.\ \eqref{eq:sing-bound-1})
$$
\sigma_{\kappa}(\bfM) \dotleq \rho^{\frac{1}{2}-\frac{rT}{2\kappa}-\frac{1}{2}\alpha_{\nt}} \leq \rho^{\frac{1}{2}-\frac{rT}{2\kappa}}
$$
where the last inequality follows as $\alpha_{\nt} > 0$. Combined with \eqref{eq:sing-bound-1} it follows that
$$
\left[ \frac{\sigma_{\kappa}(\bfM)}{\sigma_{1}(\bfM_{1})} + 1 \right] \dotleq \rho^{\frac{1}{2}\delta} \, ,
$$
and from \eqref{eq:perturbationbound-2} and \eqref{eq:sing-bound-2} that
$$
\sigma_{i}(\bfR_{k}) \dotleq \rho^{\frac{1}{2}-\frac{rT}{2\kappa}-\frac{1}{2}\alpha^\star_{\iota_{T}(i)} + \frac{3}{2}\delta}
$$
for $i=1,\ldots,qT$. Consequently (cf.\ \eqref{eq:lm-lower}),
\begin{equation} \label{eq:R-lower-bound-t}
\frac{2 \xi }{\sqrt{2k} \sigma_{i}(\bfR_{k})} \dotgeq \rho^{\frac{rT}{2\kappa}+\frac{1}{2}\alpha_{\iota_{T}(i)}^\star -\frac{1}{2} - \frac{3}{2}\delta} \, ,
\end{equation}
given that $\delta$ is sufficiently small. Further, as
$$
\frac{rT}{2\kappa}+\frac{1}{2}\alpha_{\iota_{T}(i)}^\star -\frac{1}{2} = \frac{1}{2} \left( \frac{r}{\nt} - 1 + \alpha_{\iota_{T}(i)}^\star \right) > 0
$$
for $i=1,\ldots,k$ where $k=2qT$ by the condition for $q$ in \eqref{eq:pos-constraint}, it follows that the lower bound in \eqref{eq:R-lower-bound-t} tends to infinity with increasing $\rho$ provided that $\delta$ is small, and we may conclude that
\begin{equation} \label{eq:R-lower-bound}
\left[
\frac{2 \xi }{\sqrt{2k} \sigma_{i}(\bfR_{k})} - \sqrt{2k} \right]^2 \dotgeq \rho^{\frac{rT}{\kappa}+\alpha_{\iota_{T}(i)}^\star - 1 - 3\delta} > 0
\end{equation}
where the last inequality holds, again, provided $\delta$ is small.

Combining \eqref{eq:lm-lower} in Lemma \ref{lm:upper-bound} and \eqref{eq:R-lower-bound}, and making the real valued expansion as we did for Theorem \ref{thrm:upper-bound}, yields a lower bound on the number of nodes visited by the sphere decoder in layer $k=qT$ given by
\begin{equation} \label{eq:nodes-lower-bound-1}
N_{k} \dotgeq \prod_{i=1}^{k} \rho^{\frac{rT}{\kappa}+\alpha_{\iota_{T}(i)}^\star -1 - 3\delta} = \rho^{\upsilon - 3k\delta}
\end{equation}
where
\begin{align}
\upsilon \defeq \, & \sum_{i=1}^{k} \frac{rT}{\kappa}+\alpha_{\iota_{T}(i)}^\star -1 \nonumber \\
= \, & \sum_{i=1}^q T \left( \frac{r}{\nt}+\alpha_{i}^\star-1 \right) \, .
\label{eq:upsilon}
\end{align}
By noting that
$$
0 \leq \frac{r}{\nt}+\alpha_{i}^\star-1 \leq \frac{r}{\nt}
$$
for $i=1,\ldots,q$ by the assumption that $\alpha^\star_i \leq 0$ and the definition of $q$, it follows that
\begin{equation} \label{eq:up-saturate}
T \left( \frac{r}{\nt}+\alpha_{i}^\star-1 \right) =  T \min\!
 \left( \frac{r}{\nt} - 1 + \alpha^\star_{i}\, , \, \frac{r}{\nt} \right)^{\!+}
\end{equation}
for $i=1,\ldots,q$.  Further, as
$$
\frac{r}{\nt}+\alpha_{i}^\star-1 \leq 0
$$
for $i > q$, also by the definition of $q$, the right hand side of \eqref{eq:up-saturate} is equal to $0$ for $i > q$. Thus, it follows that
$$
\upsilon = \sum_{i=1}^\nt T \min\!
 \left( \frac{r}{\nt} - 1 + \alpha^\star_{i}\, , \, \frac{r}{\nt} \right)^{\!+} = \bar{c}(r)
$$
where the last equality follows due to the optimality of $\bfalpha^\star$ in \eqref{eq:optimization}, and we then obtain from \eqref{eq:nodes-lower-bound-1} that
$$
N \geq N_{k} \geq \rho^{\bar{c}(r)-3k\delta} \, ,
$$
given that $\rho$ is sufficiently large and that $\delta > 0$ is small. However, as $\delta > 0$ can be chosen arbitrarily small it is concluded that \eqref{eq:conditions} represents sufficient conditions under which the number of nodes visited is arbitrarily close to the upper bound of $\rho^{\bar{c}(r)}$ given by Theorem \ref{thrm:upper-bound}.

\subsection{Probabilities} \label{sec:probability}

We now turn to the probability that the conditions imposed by \eqref{eq:conditions} are simultaneously satisfied. The events in \eqref{eq:conditions} are independent\footnote{The independence of $\Omega_{1}$ and $\Omega_{2}$ follows by the i.i.d.\ Rayleigh assumption on $\bfH$, which make the singular values and singular vectors of $\bfH\hr\bfH$ independent \cite{TV:04}.}. As \eqref{eq:conditions} imply $N \geq \rho^{\bar{c}(r)-\frac{3}{2}k\delta}$ it follows that
$$
\prob{N \geq \rho^{\bar{c}(r)-3k\delta}} \geq \prod_{i=1}^4 \prob{\Omega_{i}} \, ,
$$
given that $\rho$ is sufficiently large.

The assumption made in Lemma~\ref{lm:lower-bound}, i.e., condition \eqref{eq:rank-condition}, guarantees that
$$
\sigma_{1}((\bfI \kron \bfU_{p}\hr) \bfG_{|p}) > 0
$$
for some $\bfU_{p}$. However, by the continuity of singular values \cite{HJ:85} it follows for sufficiently small $u > 0$ (cf.\ \eqref{eq:condition-u}) that $\prob{\Omega_{2}} > 0$, which implies $\prob{\Omega_{2}} \doteq \rho^0$ as $\Omega_{2}$ is independent of $\rho$. The same is true for $\Omega_{3}$, i.e., $\prob{\Omega_{3}} \doteq \rho^0$. It may also be shown that $\prob{\Omega_{4}}$ converges to a strictly positive limit\footnote{This is provided that $r > 0$ in which case the the subset of the constellation defined by $\Omega_4$ contains an asymptotically deterministic and strictly positive fraction of the full constellation, cf.\ the proof of Lemma 1 in \cite{JE:10}. When $r=0$ the statement that $\bar{c}(r)$ is tight is trivial as $\bar{c}(0) = 0$.} and that therefore $\prob{\Omega_{4}} \doteq \rho^0$. It follows that
$$
\tprob{N \geq \rho^{\bar{c}(r)-3k\delta}} \dotgeq \prob{\Omega_{1}} \, .
$$

The probability of $\Omega_{1}$ may again be assessed by using large deviation techniques as in \cite{ZT:03}. In particular, it is noted that the condition imposed by $\Omega_{1}$ (cf.\ \eqref{eq:condition-alpha}) specifies an open set of admissible $\bfalpha$. Applying \eqref{eq:ld-open-bound} and \eqref{eq:rate-function} yields
\begin{align}
-\lim_{\rho \rightarrow \infty}\frac{\log \prob{\Omega_{1}}}{\log \rho} \leq & \, \sum_{i=1}^{q} (\nr-\nt+2i-1) (\alpha_{i}^{\star}-2\delta) \nonumber \\
\leq & \, d(r) - 2(\nr-\nt+q) q \delta < d(r) \label{eq:p-lower-bound-final} \, ,
\end{align}
where the second inequality follows from \eqref{eq:optimization-div} and the feasibility of $\bfalpha^\star$. Thus,
\begin{equation} \label{eq:lower-bound-2}
- \lim_{\rho \rightarrow \infty} \frac{\log \tprob{N \geq \rho^{\bar{c}(r)-3k\delta}}} {\log \rho} < d(r) \, .
\end{equation}
By the definition of the SD complexity exponent $c(r)$ (cf.\ \eqref{eq:complexity-exponent}) it follows by \eqref{eq:lower-bound-2} that $c(r) \geq \bar{c}(r)-3k\delta$. As the bound holds for arbitrarily small $\delta > 0$, it follows that $c(r) = \bar{c}(r)$, establishing the tightness of \eqref{eq:optimization} and Lemma \ref{thrm:upper-bound}.

\subsection{The extension to adaptive radius updates} \label{app:radius-updates}

The derivations above make the assumption that the search radius $\xi$ is a non-random function of $\rho$ that satisfies $\xi \doteq \rho^0$. It is thus natural to ask if the SD complexity exponent could potentially be improved by choosing $\xi$ adaptively based on the problem data $\bfH$ and $\bfY$, as is done when using, e.g., the Schnorr-Euchner SD algorithm implementation \cite{AEV:02,DGC:03}. However, we will show here that it can not, and therefore that the assumption of a non-adaptive radius is made without loss of generality. The argument is similar to the one in \cite{JO:05}.

To this end, we note that even if $\xi$ is adaptively chosen, it cannot be chosen smaller than the distance to the closest codeword, i.e., the (square root of the) minimum metric in \eqref{eq:sd-metric}, because otherwise no codeword will be chosen by the search.  As was already argued in Section~\ref{sec:search-radius}, the distance to the \emph{transmitted} codeword is $\|\bfQ\hr\bfw\|$ and $\tprob{\|\bfQ\hr\bfw\| \geq \epsilon}$ can be made arbitrarily close to one by appropriately choosing $\epsilon > 0$. Consequently, whenever the transmitted codeword $\bfs$ is the ML decision, i.e., yields the minimum metric in \eqref{eq:sd-metric}, we could use $\xi \geq \epsilon$ where $\epsilon \doteq \rho^0$ as an (arbitrarily likely) probabilistic lower bound on an adaptive search radius $\xi$ in the proof of the lower bounds on the complexity (e.g., in \eqref{eq:R-lower-bound}) and obtain the same SNR exponents in these bounds.

Thus, to complete the argument we must only rule out the possibility that the probability that $\bfs$ yields the minimum metric in \eqref{eq:sd-metric} under the conditions of $\Omega$ imposed in \eqref{eq:conditions} is small, as the lower bound is derived explicitly under $\Omega$. To this end, assume that it is false, i.e., that $\tprob{\bfs_\ml \neq \bfs | \Omega} \geq \epsilon$ for any (fixed and SNR independent) $\epsilon > 0$. In this case we could lower bound the error probability of the ML decoder according to (cf.\ \eqref{eq:p-lower-bound-final})
$$
\tprob{\bfs_\ml \neq \bfs} \geq \tprob{\bfs_\ml \neq \bfs | \Omega} \prob{\Omega} \geq \epsilon \prob{\Omega} \dotge \rho^{-d(r)} \, ,
$$
which would violate the definition of $d(r)$ as the diversity order of the ML decoder. Consequently, at sufficiently high SNR it must hold that $\tprob{\bfs_\ml = \bfs | \Omega} \geq 1-\epsilon$ for any $\epsilon > 0$. We can thus choose $\epsilon > 0$ such that, under $\Omega$, it follows that $\bfs_\ml = \bfs$ and $\|\bfQ\hr\bfw\| \geq \epsilon$ with arbitrary high probability, implying that $\xi_\ml > \epsilon$ where $\xi_\ml^2$ is the minimum metric in \eqref{eq:sd-metric}. In other words, there is some $\epsilon > 0$ for which
$$
\tprob{\Omega \cup \{ \xi_\ml \geq \epsilon \} } \dotge \rho^{-d(r)} \, .
$$
Completing the proof of Lemma~\ref{lm:lower-bound} with $\epsilon \doteq \rho^0$ in place of $\xi$ (as $\xi \geq \xi_\ml \geq \epsilon$ throughout the search) proves that $c(r) = \bar{c}(r)$ under \eqref{eq:rank-condition} also if we allow for SD implementations that adaptively choose and update the search radius $\xi$. Finally, it should be noted that what is shown here is not that adaptively choosing the search radius cannot reduce complexity -- it does -- but only that this reduction is not significant enough to reduce the complexity exponent.

\section{Proof of Lemma 3} \label{app:perturbation}

Consider the matrix, $\underline{\bfA}$, given by
$$
\underline{\bfA} \defeq \bfU \underline{\bfSigma} \bfV\hr \, ,
$$
where $\underline{\bfSigma} = \Diag(0,\ldots,0,\sigma_{i+1}(\bfA),\ldots,\sigma_{n}(\bfA))$, and where $\bfU$ and $\bfV$ denote the right and left singular vector of $\bfA$ respectively. Partition $\underline{\bfA} \in \complex^{m \times n}$ according to
$$
\underline{\bfA} = \begin{bmatrix} \underline{\bfA}_{1} & \underline{\bfA}_{2} \end{bmatrix}
$$
where $\underline{\bfA}_{1} \in \complex^{m \times n-k}$ and $\underline{\bfA}_{2} \in \complex^{m \times k}$.

By the nature of the QR decomposition, it holds that
$$
\bfP \defeq \bfPi_{\bfA_{1}}^{\perp} \bfA_{2} = \bfQ_{2} \bfR_{22}
$$
where $\bfPi_{\bfA_{1}}^{\perp}$ denotes the projection onto the orthogonal complement of the range of $\bfA_{1}$ (i.e.\ the nullspace of $\bfA_{1}\hr$). Additionally, let
$$
\underline{\bfP} \defeq \bfPi_{\underline{\bfA}_{1}}^{\perp} \underline{\bfA}_{2} \, .
$$
As $\bfQ_{2}$ is a unitary matrix it follows that
$$
 \sigma_{i}(\bfR_{22}) = \sigma_{i}(\bfP) \, .
$$
In what follows, we will consider the singular values of $\bfP$ in order to establish the lemma. To this end, we will make use of two results due to Weyl and Stewart.  For a modern proof of Theorem \ref{thrm:weyl}, see e.g.\ \cite[Corollary 7.3.8]{HJ:85}. The statement in Theorem \ref{thrm:stewart} follows by combining \cite[Theorem 2.3]{Ste:77} and \cite[Theorem 2.4]{Ste:77}.  In the following, $\|\bfB\|=\sigma_{\max}(\bfB)$ denotes the spectral matrix norm.

\begin{theorem}[Weyl] \label{thrm:weyl}
For arbitrary $\bfB,\bfC \in \complex^{p \times q}$ it holds that
\begin{equation} \label{eq:weyl}
| \sigma_{i}(\bfB) - \sigma_{i}(\bfC) | \leq \| \bfB - \bfC \| \, .
\end{equation}
\end{theorem}

\begin{theorem}[Stewart] \label{thrm:stewart}
For $\bfB,\bfC \in \complex^{p \times q}$ such that $\rank(\bfB) = \rank(\bfC)$,
\begin{equation} \label{eq:stewart}
\| \bfPi_{\bfB}^\perp - \bfPi_{\bfC}^\perp \| \leq \min( \| \bfB^\dagger \| \, , \, \| \bfC^\dagger \| ) \, \|\bfB-\bfC\| \, ,
\end{equation}
where $(\cdot)^\dagger$ denotes the Moore-Penrose pseudo inverse \cite{HJ:85}.
\end{theorem}

By noting that
\begin{equation} \label{eq:subblockbound}
\| \bfA_{l} - \underline{\bfA}_{l}\| \leq \|\bfA - \underline{\bfA} \| = \sigma_{i}(\bfA) \, ,
\end{equation}
for $l=1,2$ and using the assumption that $\sigma_{1}(\bfA_{1}) > \sigma_{i}(\bfA)$ it follows by Theorem \ref{thrm:weyl} that
$$
\sigma_{1}(\underline{\bfA}_{1}) \geq \sigma_{1}(\bfA_{1}) - \| \bfA_{1} - \underline{\bfA}_{1} \| \geq \sigma_{1}(\bfA_{1}) - \sigma_{i}(\bfA) > 0
$$
implying that $\underline{\bfA}_{1}$ is full rank. As $\sigma_{1}(\bfA_{1}) > 0$ is directly implied by $\sigma_{1}(\bfA_{1}) > \sigma_{i}(\bfA)$ it follows that $\rank(\underline{\bfA}_{1}) = \rank(\bfA_{1})$ which makes Theorem \ref{thrm:stewart} applicable to $\bfPi_{\bfA_{1}}^\perp - \bfPi_{\underline{\bfA}_{1}}^\perp$. As
\begin{align*}
\bfP - \underline{\bfP} & = \bfPi_{\bfA_{1}}^{\perp} \bfA_{2} - \bfPi_{\underline{\bfA}_{1}}^{\perp} \underline{\bfA}_{2} \\
& = (\bfPi_{\bfA_{1}}^\perp - \bfPi_{\underline{\bfA}_{1}}^\perp) \underline{\bfA}_{2} + \bfPi_{\bfA_{1}}^\perp(\bfA_{2}-\underline{\bfA}_{2})
\end{align*}
it follows that
$$
\| \bfP - \underline{\bfP} \| \leq \| \bfA_{1}^\dagger \| \| \bfA_{1}-\underline{\bfA}_{1} \| \| \underline{\bfA}_{2} \| + \| \bfA_{2} - \underline{\bfA}_{2} \| \, ,
$$
where we used Theorem \ref{thrm:stewart} and the fact that $\|\bfB \bfC\| \leq \| \bfB \| \| \bfC \|$ and $\|\bfPi_{\bfB}^\perp\| \leq 1$ \cite{HJ:85}. By noting that $\| \bfA_{1}^\dagger \| = 1/\sigma_{1}(\bfA_{1})$, that $\| \underline{\bfA}_{2}\| \leq \| \underline{\bfA}\| = \sigma_{n}(\bfA)$, that $\| \bfA_{1} - \underline{\bfA}_{1} \| \leq \sigma_{i}(\bfA)$ and that $\| \bfA_{2} - \underline{\bfA}_{2} \| \leq \sigma_{i}(\bfA)$ (cf.\ \eqref{eq:subblockbound}), it follows that
\begin{equation} \label{eq:normbound}
\mu \defeq \left[ \frac{\sigma_{n}(\bfA)}{\sigma_{1}(\bfA_{1})} + 1 \right] \sigma_{i}(\bfA) \geq \| \bfP - \underline{\bfP} \|.
\end{equation}
By again applying Theorem \ref{thrm:weyl} to \eqref{eq:normbound} it follows that $\sigma_{i}(\bfP) \leq \sigma_{i}(\underline{\bfP}) + \mu$. Note however that
$$
\rank(\underline{\bfA}) = \rank(\underline{\bfA}_{1}) + \rank(\underline{\bfP})
$$
where $\underline{\bfP} \defeq \bfPi_{\underline{\bfA}_{1}}^\perp \underline{\bfA}_{2} \in \complex^{m \times k}$. As $\rank(\underline{\bfA}_{1}) = n-k$ and $\rank(\underline{\bfA}) \leq n-i$ it follows that
$$
\rank(\underline{\bfP}) \leq k-i
$$
and $\sigma_{i}(\underline{\bfP}) = 0$. Thus, $\sigma_{i}(\bfR_{22}) = \sigma_{i}(\bfP) \leq \mu$ establishing the lemma.

\section{Proof of Theorem 7} \label{app:nvd}
Let $\underline{\Xset}$ be the un-normalized extended codebook corresponding to the un-normalized lattice points $\bfG\constme_\infty$, i.e., where $\Xset \subseteq \theta \underline{\Xset}$.  A space-time code is said to satisfy the non-vanishing determinant (NVD) condition if \cite{OBV:07}
\begin{equation} \label{eq:nvd}
\inf_{\underline{\bfX} \in \underline{\Xset} \backslash \bfZero} |\underline{\bfX}| > 0 \, ,
\end{equation}
i.e., if there are no non-zero un-normalized (difference) codewords with arbitrarily small determinants. The proof of the theorem draws from the well known fact that the NVD condition is a necessary condition for achieving approximate universality, and divides the problem into a few (exhaustive) cases where either the condition in Lemma~\ref{lm:lower-bound} is shown to hold, or the NVD property is shown to be violated, thus eliminating the possibility of NVD codes that would violate the rank condition in Lemma~\ref{lm:lower-bound}. To this end, consider a partitioning of the $4 \times 4$ generator matrix according to
$$
\bfG =
\begin{bmatrix}
\bfG_{11} & \bfG_{12} \\
\bfG_{21} & \bfG_{22}
\end{bmatrix}
$$
where $\bfG_{ij} \in \complex^{2 \times 2}$. First of all, let us note that the case where $p=2$ is trivially satisfied as $\bfG$ is full rank, i.e., the matrix
$$
(\bfI_2 \kron \bfU_2\hr)\bfG
$$
is full rank for any unitary $\bfU_2 \in \complex^{2 \times 2}$. We can thus restrict attention to the case of $p=1$ and consider the rank of
\begin{equation} \label{eq:2x2rank}
(\bfI_2 \kron \bfu\hr)\bfG_{|1} = \begin{bmatrix}
\bfu\hr\bfG_{11} \\
\bfu\hr\bfG_{21}
\end{bmatrix} \in \complex^{2 \times 2}
\end{equation}
where $\bfu = \bfU_1 \in \complex^{2 \times 1}$. In the cases where the NVD property is shown to not hold, it is sufficient to consider non-zero (unnormalized) codewords of the form
$$
\underline{\bfx} = \begin{bmatrix}
\bfG_{11} & \bfG_{12} \\
\bfG_{21} & \bfG_{22}
\end{bmatrix} \begin{bmatrix}
\bfs_{1} \\
\bfs_{2}
\end{bmatrix}
$$
where $\bfs_{1} \in \constme_\infty^2$, and $\bfs_1 \neq 0$, and where $\bfs_2 = 0$. The (un-normalized) codewords in matrix form are in this case given by $\underline{\bfX} = [ \bfG_{11} \bfs_1 \; \bfG_{21} \bfs_1]$ and we have $\bfu\hr\underline{\bfX} = [ \bfu\hr\bfG_{11}\bfs_1 \; \bfu\hr\bfG_{21} \bfs_1 ]$. All codewords discussed in what follows are assumed to have this structure.

We will now consider different cases depending on the rank of $\bfG_{11}$ and $\bfG_{21}$. However, as it is straightforward to see that $\underline{\bfX}$ has zero determinant (for any $\bfs_1$) if either $\bfG_{11} = \bfZero$ or $\bfG_{21} = \bfZero$, the cases that need consideration are those when the rank of both $\bfG_{11}$ and $\bfG_{21}$ is equal to one (case a), when the rank of both $\bfG_{11}$ and $\bfG_{21}$ is equal to two (case b), and when the rank or either $\bfG_{11}$ or $\bfG_{21}$ is equal to one and the rank of the other is equal to two (case c).

\subsection{Case a}

Consider the case where the rank of both $\bfG_{11}$ and $\bfG_{21}$ is one, i.e., where
$\bfG_{11} = \bfb_1\bfa_1\hr$ and $\bfG_{21} = \bfb_2\bfa_2\hr$. If $\bfa_1$ and $\bfa_2$ are not linearly dependent, the condition in \eqref{eq:2x2rank} is satisfied for any $\bfu$ such that $\bfu\hr\bfb_1 \neq 0$ and $\bfu\hr\bfb_2 \neq 0$, and we can thus restrict attention to the case where $\bfa_1$ and $\bfa_2$ are linearly dependent. Here, we may without loss of generality assume that $\bfa_1 = \bfa_2 = \bfa$ by absorbing any complex scalars into $\bfb_1$ and $\bfb_2$.

Note however that given any $\epsilon > 0$ we can always find a point $\bfs_1 \in \constme_\infty^2$, where $\bfs_1 \neq \bfZero$ such that\footnote{Since $\bfa\hr\constme_\infty^2$ is the projection of the two dimensional Gaussian integer lattice onto a one dimensional space, one can also view the statement as an application of Dirichlet's box principle, cf.\ \cite{TK:10}. In general, the problem of finding non-zero integer vectors that are approximately orthogonal to a given vector is known as approximate integer relations (IRs), and is related to simultaneous Diophantine approximations \cite{Cla97}.} $\|\bfa\hr\bfs_1\|^2 < \epsilon$. For any such $\bfs_1$ it follows that (cf.\ \cite[Theorem 7.3.10]{HJ:85})
$$
\sigma_{\max}^2(\underline{\bfX}) = \max_{\bfu \in \complex^{2} : \|\bfu\|=1} \|\bfu\hr\underline{\bfX}\|^2 \leq (\|\bfb_1\|^2 + \|\bfb_2\|^2) \epsilon \, ,
$$
i.e., the maximal singular value of $\underline{\bfX}$ can be made arbitrarily small. However, this violates the assumed NVD property of the code as a small maximal singular value implies a small determinant, and concludes case a.

\subsection{Case b}

When $\bfG_{11}$ and $\bfG_{21}$ are full rank we can always find a vector $\bfu$ such that $\bfu\hr\bfG_{11}$ and $\bfu\hr\bfG_{21}$ are linearly independent (thus satisfying the condition of Lemma~\ref{lm:lower-bound}) unless $\bfG_{11}$ and $\bfG_{21}$ are linearly dependent, i.e., when $\bfG_{11} = a \bfG_{21}$ for some $a \in \complex$. However, in this case we have that $\bfG_{11} \bfs_1 = a \bfG_{21}\bfs_1$ for any $\bfs_1$ which implies that the columns of $\underline{\bfX}$ are linearly dependent, and the rank of $\underline{\bfX}$ is zero. This concludes case b.

\subsection{Case c}

In this case we may assume that $\bfG_{11} = \bfb_1\bfa_1\hr$ has rank one and $\bfG_{21}$ has rank two (the opposite case is handled equivalently). Here, as both the set of $\bfu$ for which $\bfu\hr\bfb_1 = \bfZero$ and where $\bfu\hr\bfG_{21}$ is linearly dependent of $\bfa_1\hr$ have zero measure, we may pick $\bfu$ such that $\bfu\hr\bfb_1 \neq \bfZero$ and such that $\bfu\hr\bfG_{21}$ is linearly independent  of $\bfa_1\hr$, thus satisfying the conditions of Lemma~\ref{lm:lower-bound}. This concludes the proof of Theorem~\ref{thrm:nvd}.

\end{appendices}

\end{document}